\documentclass[%
reprint,
superscriptaddress,
amsmath,amssymb,
aps,
prb,
floatfix,
]{revtex4-2}
\usepackage{graphicx}
\usepackage{dcolumn}
\usepackage{bm}
\usepackage{chemformula}
\usepackage{hyperref}
\usepackage{tabularx}
\usepackage{amsmath}
\usepackage{float}

\begin{document}

    \title{Probing ice-rule-breaking transition in \ch{Dy_2Ti_2O_7} thin film by proximitized transport and magnetic torque}
	
	\author{Chengkun Xing}
	\affiliation{Department of Physics and Astronomy, University of Tennessee, Knoxville, TN 37996, USA}
	\author{Han Zhang}
	\affiliation{Department of Physics and Astronomy, University of Tennessee, Knoxville, TN 37996, USA}
	\author{Kyle Noordhoek}
	\affiliation{Department of Physics and Astronomy, University of Tennessee, Knoxville, TN 37996, USA}
        \author{Guoxin Zheng}
	\affiliation{Department of Physics, University of Michigan, Ann Arbor, Michigan 48109, USA}
        \author{Kuan-Wen Chen}
	\affiliation{Department of Physics, University of Michigan, Ann Arbor, Michigan 48109, USA}
	\author{Lukas Horák}
	\affiliation{Charles University, Prague, Czechia,11636}
	\author{Yan Xin}
	\affiliation{National High Magnetic Field Laboratory, Florida State University, Tallahassee, FL 32310, USA}
	\author{Eun Sang Choi}
	\affiliation{National High Magnetic Field Laboratory, Florida State University, Tallahassee, FL 32310, USA}
        \author{Lu Li}
	\affiliation{Department of Physics, University of Michigan, Ann Arbor, Michigan 48109, USA}
	\author{Haidong Zhou}
	\email{hzhou10@utk.edu}
	\affiliation{Department of Physics and Astronomy, University of Tennessee, Knoxville, TN 37996, USA}
	\author{Jian Liu}
	\email{jianliu@utk.edu}
	\affiliation{Department of Physics and Astronomy, University of Tennessee, Knoxville, TN 37996, USA}
	
	\date{\today}%
	
	\begin{abstract}
		{
    While the spin ice state of bulk pyrochlores such as \ch{Dy2Ti2O7} and \ch{Ho2Ti2O7} has been extensively studied in the last several decades due to its unique degenerate ground state and emergent monopole excitation, whether it survives in the thin-film form remains a mystery. The limited volume of thin-film sample makes it challenging to study the intrinsic magnetic properties. Here, we synthesized 18nm-thick \ch{Dy2Ti2O7} thin film on YSZ \textcolor{black}{(Yttria-stabilized Zirconia with 9.5 mol\% \ch{Y2O3})} substrate and capped it by a thin conductive \ch{Bi2Ir2O7} layer, and performed the proximitized magnetoresistance measurements. Our study found that the ice-rule-breaking phase transition survives but with a modified effective nearest-neighbor interaction \textcolor{black}{($\rm{J_{eff}}=$ 1.054 K)} and distorted Ising spin axes \textcolor{black}{($\rm{\epsilon}=+0.051)$} compared to the bulk crystal. The results are supported by the simultaneously measured capacitive torque magnetometry. Our study demonstrates that proximitized transport is an effective tool for thin films of insulating frustrated magnets.
		}
	\end{abstract}
	\maketitle

    Geometrical frustration is an important driving force for novel emergent magnetic states. One of the most important prototypes is the classical spin ice found in rare earth pyrochlores, such as \ch{Dy2Ti2O7} (DTO) and \ch{Ho2Ti2O7} (HTO) \cite{gardner2010magnetic,harris1997geometrical,isakov2005spin,siddharthan1999ising,bramwell1998frustration,bramwell2020history,snyder2001spin}. It is a magnetic analog of water ice \cite{pauling1935structure} and exhibits emergent magnetic monopole excitation \cite{castelnovo2008magnetic,fennell2009magnetic,jaubert2011magnetic,jaubert2009signature,morris2009dirac,khomskii2012electric}. The basic building block is the tetrahedron of the Ising spins \cite{tomasello2015single,jana2002estimation} that point either in or out of the tetrahedron and must settle in one of the six degenerate 2-in-2-out (2:2) spin configurations \cite{bramwell2001spin,ruff2005finite,den2000dipolar,bramwell2001spinscience}. The degeneracy rapidly grows as the (2:2) tetrahedra connect to each other according to the ice rule within the pyrochlore structure, giving rise to the non-zero residual entropy  \cite{ramirez1999zero}. More interestingly, flipping a spin necessarily breaks the ice rule and effectively creates a pair of magnetic monopole and antimonopole by turning two (2:2) tetrahedra into the 1-in-3-out and 3-in-1-out configurations, respectively \cite{sakakibara2003observation,fukazawa2002magnetic,petrenko2003magnetization}. When such spin flip is induced everywhere in the lattice by a magnetic field along a $\textlangle111\textrangle$ axis, the monopole/antimonopole density increases drastically as they condensate into the magnetic charge-ordered state, i.e., the so-called saturated ice (3:1) state. This ice-rule-breaking transition is thus described as the liquid-gas transition of the magnetic monopoles. \cite{hiroi2003specific,higashinaka2004low,aoki2004magnetocaloric,grigera2015intermediate}.

    Although rich spin ice physics has been extensively studied in bulk crystals during the last three decades, epitaxial engineering has recently been considered as a new route to tuning the spin ice behavior due to the possibility of controlling, for instance, the dimensionality \cite{bergmann1984weak} and/or strain effect \cite{pandey2022controllable}. However, experimental results on spin ice pyrochlore thin films have been relatively limited, and the conclusions are far from clear \cite{barry2019modification,leusink2014thin,bovo2014restoration,bovo2019phase, wen2021epitaxial,miao2020two,lantagne2018spin,jaubert2017spin}. In particular, two studies showed that the magnetic anisotropy and magnetization of thick HTO films are similar to the HTO crystal \cite{leusink2014thin,barry2019modification}, which was measured above 1.8 K and well above the temperatures of spin-ice behavior \cite{bramwell2001spin,hiroi2003specific}. On the other hand, two other studies suggested that the Pauling entropy is released in DTO films based on heat capacity and that a spin-ordered state may exist \cite{bovo2014restoration,bovo2019phase}, potentially due to the strain effect according to Monte Carlo simulations \cite{jaubert2017spin}. However, the exact magnetic structure of the ordered state is not yet clear. While it would be highly interesting if the spin-ice state is completely suppressed in thin films and/or new states emerge, the foremost challenge is to establish benchmark films with the spin-ice states that are comparable to bulk crystals. However, the limited film studies highlight the general experimental challenge in characterizing thin films of insulating frustrated magnets due to the small sample volume, the presence of the substrate, and the requirement of reaching ultralow temperatures, all of which together hinder the use of conventional bulk techniques, such as AC susceptibility and neutron scattering. Methods that are applicable to both bulk and film, even in the ultrathin limit, are crucial for overcoming this barrier. Our recent work on epitaxial \ch{Bi2Ir2O7} (BIO) thin films deposited on DTO single crystal substrates has demonstrated that BIO resistivity is sensitive to the ice-rule-breaking transition of bulk DTO through a resistance anomaly that faithfully tracks the transition \cite{zhang2023anomalous}. This method utilizes BIO as a non-magnetic pyrochlore metal \cite{qi2012strong,chu2019possible,yang2018stoichiometry} that can be grown epitaxially on DTO. Such a proximitized electronic transport offers a new route to characterizing insulating frustrated magnets and is technically quite applicable even for thin-film samples but needs to be demonstrated.

\begin{figure}[t]
    \centering{\includegraphics[clip,width=8cm]{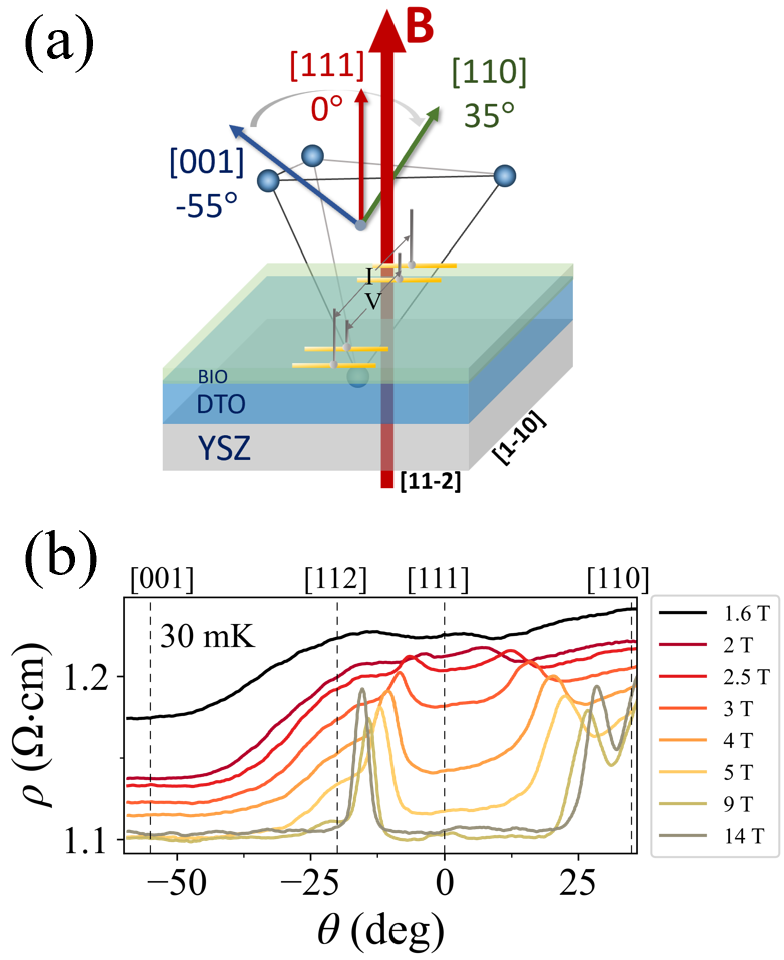}}
    \caption{(a): Schematic diagram of the BIO/DTO bilayer on the YSZ substrate and the proximitized transport measurement configuration. The current is along [1-10] axis, and the magnetic field is \textcolor{black}{in (1-10) plane and} always perpendicular to the current. (b): The measured BIO resistivity as a function of the field angle in (1-10) plane at different applied magnetic field at 0.03 K. 
    }
    \label{fig1}
\end{figure}

    Here we apply this method of proximitized transport to a (111)-oriented DTO thin film by measuring the resistance of a BIO epilayer (Fig. \ref{fig1}(a)) and demonstrate the existence of the ice-rule breaking transition. To independently validate the results, we simultaneously measure the capacitive torque magnetometry (CTM) which is sensitive to the bulk of the DTO thin film. The results unambiguously show close correspondence between the two measurements. Proximitized transport thus allows comparing film samples with the bulk crystals of classical spin ice and is promising for studying epitaxially engineered insulating frustrated magnets. Detailed information about the growth of the BIO/DTO bilayer on the YSZ substrate can be found in End Matter.

Fig. \ref{fig1}(b) shows the magnetic field angle scan of the BIO resistivity at different field values at 0.03 K by rotating the field around [111]. Since the field was applied in the (1-10) plane (Fig. \ref{fig1}(a)), the angle scan allowed the field to cover the three high-symmetry axes from [001] to [111] and to [110]. We defined the [111] direction as the zero angle so that [110] and [001] are at $\theta=$ 35$^{o}$ and -55$^{o}$, respectively. The current was along [1-10] and thus always normal to the field. One can see that the angle dependence is quite flat at 14 T, except for a couple of sharp peaks, one on the positive side and the other one on the negative side of [111]. As the field decreases, these peaks shift toward [111] and broaden. Although the overall angle dependence is not as flat at 5 T and below,  one can still see that the heights of the two peaks decrease as they shift toward [111] until they disappear around 2 T. The observed shift of the peaks is reminiscent of the field-angle phase diagram of classical spin ice where the critical angle of the transition between (3:1) and (2:2) decreases on both sides of [111] when the field decreases \cite{sato2007field, grigera2015intermediate}. In the high-field limit, the critical angle should approach [112] and [110] for the negative and positive side, respectively, which is also consistent with the way the observed peaks shift with increasing field. These proximitized transport results point to the existence of the ice-rule-breaking transition in the DTO film similar to DTO bulk crystal.

\begin{figure}[p]
    \centering{\includegraphics[clip,width=8.5cm]{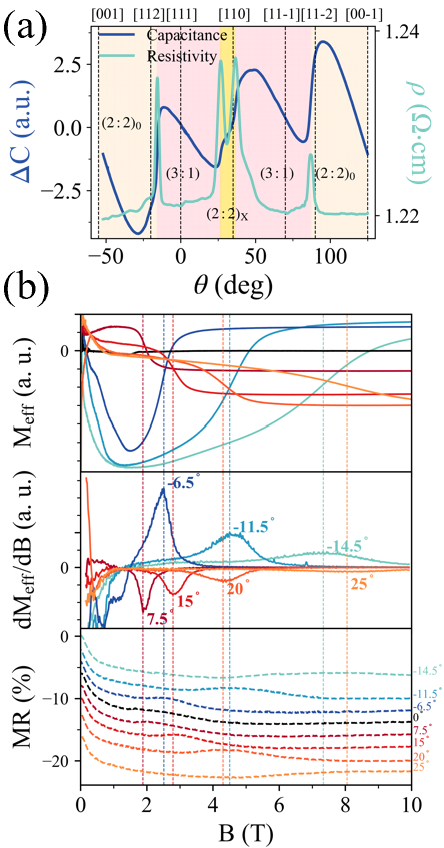}}
    \caption{(a):  The measured capacitance change $\rm{C(B)-C(0~T)}$ and BIO resistivity as a function of angle in (1-10) plane with a 10 T applied magnetic field at 0.03 K (blue line is capacitance change and green line is resistivity). The filled colors highlight the different phases. The phase boundary is defined at the position of the resistivity peaks. (b): The field dependence results at 0.03 K with the magnetic field along different directions around [111] axis in the (1-10) plane. The $\rm{\theta_c}$ values are denoted in the middle panel. Top panel: Effective magnetic moment vs. magnetic field. Middle panel: Derivative of effective magnetic moment vs. magnetic field. Bottom panel: Magnetoresistance (MR) vs. magnetic field (vertical offsets are applied on the curves for clarity). The black curve is the results when the magnetic field is applied along [111] axis. The vertical lines represent peaks/dips of effective magnetic moment derivatives for comparison with the MR anomaly.}
		\label{fig3}
	\end{figure}

While the proximitized transport in BIO is expected to be caused by the extra scattering of the electrons due to fluctuations of the Dy moments near the ice-rule-breaking transition \cite{zhang2023anomalous,ohno2024proximity}, it is necessary to have another independent measure that probes the bulk of the DTO film for comparison and verification. Capacitive torque magnetometry (CTM) is an effective tool for this purpose thanks to the high resolution of the cantilever deflection that measures magnetization vector by measuring the capacitance between the cantilever and the underneath gold film. The capacitance change is thus proportional to the torque generated by the sample under the magnetic field \cite{li2008torque}. We mounted the sample on the cantilever and also wired the sample for the resistance measurement so that the two independent measurements can be simultaneously carried out. Fig. \ref{fig3}(a) shows the resistivity and the capacitance change at 0.03 K as a function of $\theta$ when a 10 T field was applied and rotated again within the (1-10) plane. We set up the rotation of the CTM device to cover 180$^{o}$ from the [001], [111], [110], [11-1], and to [00-1] axis. The current was also along [1-10]. As one can see, when the field deviates from [111] toward [110] or [001], the BIO resistivity exhibited sharp spikes similar to Fig. \ref{fig1}(b). They occurred at the same angles as the sharp jumps of the torque. The same behavior occurred when the field deviates from [11-1] toward [110] or [00-1]. The sharp jumps of the torque are known as turnovers of the magnetization vector during the transition between (2:2) and (3:1) \cite{anand2022investigation}. Specifically, since the degeneracy of the (2:2) state is fully lifted in the high-field limit with field along [001], i.e., the $\rm{(2:2)_0}$ state, the transition into $\rm{(3:1)}$ occurs by flipping the $\alpha$ spin chain at the critical angle, causing the turnover. The degeneracy of the (2:2) state is partially lifted with field along [110], i.e., the $\rm{(2:2)_X}$ state. It transitions to $\rm{(3:1)}$ by flipping the $\beta$ spin chain at the critical angle. The overall angle dependence of the torque is in fact similar to HTO and characteristic of a classical spin-ice system. The result thus confirms that the torque signal is driven by the DTO film since neither BIO nor YSZ would display these features. The fact that the BIO resistivity also responds to the transition but with spikes instead of jumps confirms that the proximitized transport is not proportional to or directly correlated to the magnetization but driven by enhanced scattering from the unstable and fluctuating $\alpha$ or $\beta$ spin chain during the transition. The remarkable consistency of the critical angles between the two measurements indicates no significant difference between the bulk of the DTO film and its interfacial region near BIO.

    We further performed field scans to compare the proximitized transport and the CTM. Fig. \ref{fig3}(b) shows the results at 0.03 K at different $\theta$ around [111]. The top panel shows the effective magnetic moment $\mathrm{M_{eff}}$, defined as $\mathrm{(C(B) - C(0\ T))/B}$ where  $\mathrm{B}$ is the field. At high fields, $\mathrm{M_{eff}}$ always reaches a plateau, because the magnetization is saturated when the (3:1) state is fully stabilized. The plateau has opposite signs for positive and negative $\theta$ because the magnetization vector is always along [111] in (3:1) and hence at the opposite sides of the field. Before reaching the (3:1) state, however, the magnetization vector of the $\rm{(2 : 2)}$ state is along [110] or [001]. Therefore, the plateau is always reached by a drastic increase or decrease of $\mathrm{M_{eff}}$, similar to the reported CTM data on HTO crystal \cite{anand2022investigation}. This behavior corresponds to the turnovers of the magnetization vector during the ice-rule-breaking transition revealed in the angle scan [Fig. \ref{fig3}(a)]. Therefore, one can calculate the first-order derivative of this part of the $\mathrm{M_{eff}}$ curve (central panel), and define the peak or dip as the critical field $\mathrm{B_c}$ [Fig. \ref{fig3}(b)]. Interestingly, an anomaly arises exactly at $\mathrm{B_c}$ for all angles as well for the simultaneously measured magnetoresistance (MR) of the BIO layer (bottom panel). \textcolor{black}{In other words, the anomaly shows maximum resistivity when $\mathrm{M_{eff}}$ jumps from one side to another, similar to the angle scan.} Similar to the BIO film on a DTO crystal \cite{zhang2023anomalous}, the MR anomaly sits on top of the negative MR curve and shifts to higher fields as the field deviates from [111] because the ice-rule-breaking transition shifts to higher fields, which is now also captured by the CTM. In addition to the consistent $\mathrm{B_c}$ from both CTM and MR, one can see both the peak/dip of $\mathrm{dM_{eff}/dB}$ and the MR anomaly broaden at the same time as the field rotates away from [111]. This corresponds to an enlarged width of the transition, consistent with the fact that a larger field is needed to induce the transition. All the data clearly demonstrate that the ice-rule-breaking transition is preserved in the DTO thin film. Similar to the angle scan, the torque probes the transition by tracking the magnetization vector, whereas the MR anomaly reflects the fluctuations during the transition. 

\begin{figure*}[t!]
        \centering{\includegraphics[clip,width=18cm]{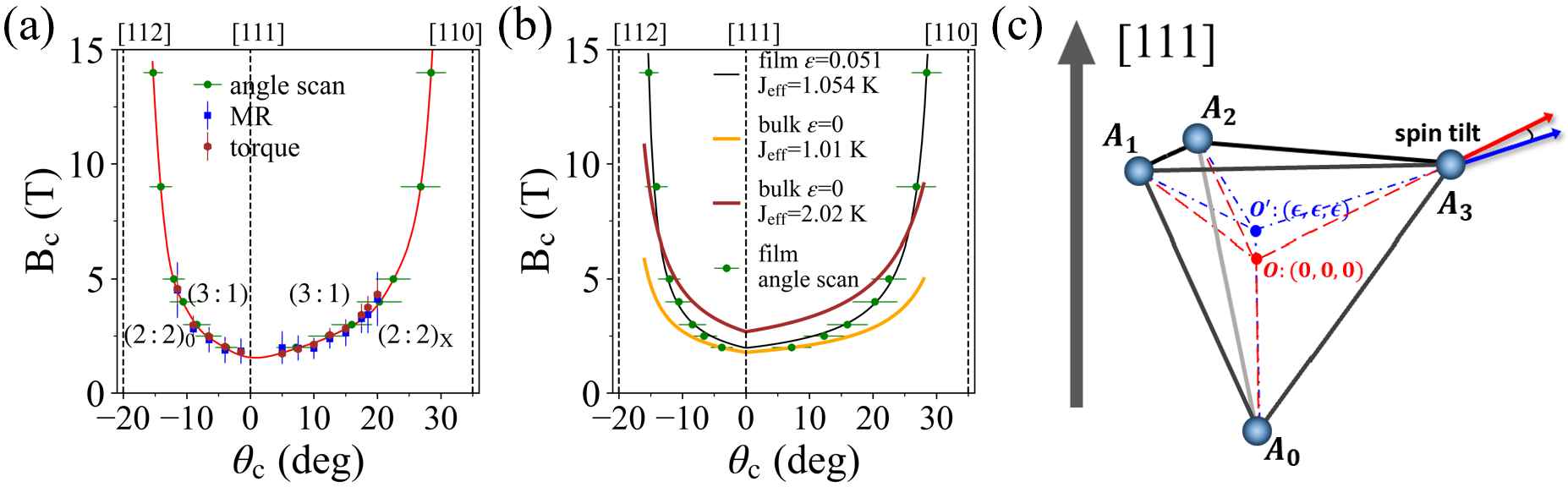}}
        \caption{(a): Green dots represent the critical angles derived from the peak of angle-dependent BIO resistivity (Fig. \ref{fig1}(b)). Blue dots represent the critical field extracted from the MR anomaly in Fig. \ref{fig3}(b) (bottom) by fitting
        exponentially modified Gaussian distribution. \textcolor{black}{Examples of the fitting are illustrated in Fig. S1 \cite{supplemental} and Fig. S2 \cite{supplemental}.} The brown dots represent the critical field which is extracted from the peaks/dips in the derivative of effective magnetic moment in Fig. \ref{fig3}(b) (middle). The red line is a guide to the eyes for illustrating the phase boundary. (b): Green dots represent the critical angles derived from the peak of angle-dependent BIO resistivity at different fields (Fig. \ref{fig1}(b)). Orange line represents the theoretical phase boundary of the DTO bulk by using Eq. \ref{eq1} with parameters $\rm{J_{eff}}=$ 1.01 K and $\mathrm{B_{dem}}=$ 0.88 T. Brown line is obtained by increasing $\rm{J_{eff}}$ by a factor of 2. Black curve represents the calculated phase boundary by using Eq. \ref{eq2} with parameters $\rm{J_{eff}}=$ 1.054 K, $\rm{\epsilon}=$ +0.051 and $\mathrm{B_{dem}}=$ 0.88 T. (c): Schematic that shows the off-center of the Ising spin axes. The coordinate of the tetrahedron corners are: $\rm{A_0(-1,-1,-1), A_1(1,-1,1), A_2(-1,1,1), A_3(1,1,-1)}$.}
        \label{fig4}
    \end{figure*}

Fig. \ref{fig4}(a) plots both the critical angle from the angle scans at different fields and the critical field from the field scans at different angles. One can see that the phase boundary is consistent between the angle scan and field scan, and the shape of the phase boundary is characteristic of the reported ones of classical spin ice \cite{anand2022investigation, sato2007field}. \textcolor{black}{The phase boundary from the CTM and the proximitized transport is consistent as well within the finite width of the transition. Since sample inhomogeneity would broaden the transition width, the consistency of the two methods shows inhomogeneity, if any, affects the bulk region and the interface region in a similar way.}

It is known that the phase boundary between (2:2) and (3:1) \cite{sato2007field,borzi2016intermediate} is directly related to the effective nearest-neighbor magnetic interaction $\rm{J_{eff}=J_{nn}+D_{nn}}$, where $\rm{J_{nn}}$ and $\rm{D_{nn}}$ are the nearest-neighbor superexchange interaction and nearest-neighbor dipolar interaction of the ``dipolar spin ice" model (DSIM) \cite{den2000dipolar,bramwell2001spinscience}. The exact field-angle dependence of the phase transition is, however, rather complicated and still debatable, especially at small $\theta$ where effects, such as further nearest-neighbor couplings and phonons, could play important roles \cite{anand2022investigation,borzi2016intermediate}. This is because the energy separation among different (2:2) configurations is small and a large degree of the degeneracy remains across the transition. In contrast, the situation is relatively straightforward at high angles where the large field lifts most, if not all, of the degeneracy and the transition occurs between $\rm{(2 : 2)_0}$ or $\rm{(2 : 2)_X}$ and $\rm{(3 : 1)}$. In this case, $\rm{B_c}$ follows an angular relation as \cite{sato2007field}:
    \begin{equation}\label{eq1}
        \scalebox{1.1}{$
    \rm{B_c} = \begin{cases}
                \rm{\frac{0.6J_{eff}}{\cos{\theta_c}+2\sqrt{2}\sin{\theta_c}}+B_{dem}} & \rm{(\theta_c < 0)}\\\\
           \rm{\frac{0.6J_{eff}}{\cos{\theta_c}-\sqrt{2}\sin{\theta_c}}+B_{dem}} & \rm{(\theta_c > 0)}
             \end{cases}
        $
    }
    \end{equation}
    where $\rm{B_{dem}}$ is the demagnetization field caused by the plate-shape of the (111)-oriented sample. This relation is derived by considering four Ising spins at the tetrahedral vertices with $\rm{J_{eff}}$ from the DSIM. (the data points from \cite{zhang2023anomalous} are consistent with the bulk phase boundary from Eq. \ref{eq1}.) However, if we simply adopt $\rm{J_{eff}}$ of bulk DTO \cite{sato2007field}, this theoretical phase boundary is clearly below $\rm{B_c}(\theta_c)$ of the DTO film, especially at high angles [orange line shown in Fig. \ref{fig4}(b)]. \textcolor{black}{Note that the proximitized transport results on DTO crystal \cite{zhang2023anomalous} are consistent with this bulk phase boundary.} While the overall upshift of the experimental phase boundary appears to indicate a larger $\rm{J_{eff}}$, simply adjusting $\rm{J_{eff}}$  [brown line shown in Fig. \ref{fig4}(b)] does not account for this inconsistency. The reason is that the deviation between the experimental and theoretical phase boundaries is asymmetric between the positive and negative sides of $\theta$, whereas $\rm{J_{eff}}$ only scales both sides equivalently according to Eq. \ref{eq1}. The fact that $\rm{J_{eff}}$ is only a scaling factor means the critical angles on the positive and negative sides under the same critical field must have a one-to-one correspondence independent of $\rm{J_{eff}}$. It is the deviation from this one-to-one correspondence that prohibits matching the theoretical phase boundary to the experimental one by simply tuning $\rm{J_{eff}}$.
    This observation suggests a geometric modification within the tetrahedron because the different angular functions of the two sides originate from flipping different spin chains during the transition. Eq. \ref{eq1} is indeed derived under the condition that the local axes of all four Ising spins intersect the center of the tetrahedron, which can be defined as the origin (0,0,0) [Fig. \ref{fig4}(c)] and dictates the angular functions. If one assumes that the crossing point shifts away from (0,0,0) along [111] to $\rm{(\epsilon,\epsilon,\epsilon)}$ and repeats the same derivation, Eq. \ref{eq1} must be modified as:
    
    \begin{equation}\label{eq2}
        \scalebox{1.1}{$
    \rm{B_c} = \begin{cases}
                \rm{\frac{0.6J_{eff}\sqrt{3\epsilon^2-2\epsilon+3}/\sqrt{3}}{\cos{\theta_c}(1-3\epsilon)+2\sqrt{2}\sin{\theta_c}}+B_{dem}} & \rm{(\theta_c < 0)}\\\\
           \rm{\frac{0.6J_{eff}\sqrt{3\epsilon^2-2\epsilon+3}/\sqrt{3}}{\cos{\theta_c}(1-3\epsilon)-\sqrt{2}\sin{\theta_c}}+B_{dem}} & \rm{(\theta_c > 0)}
             \end{cases}
        $
    }
    \end{equation}
    One can see this modified relation (black curve in Fig. \ref{fig4}(b)) now allows asymmetric deviation of the two sides from that of Eq. \ref{eq1} by changing $\rm{\epsilon}$. And $\rm{\epsilon}$ can be calculated given a pair of critical angles of the two sides without the need to know the corresponding critical field. By using the critical angles at 14 T, we obtained an $\rm{\epsilon}=+0.051 (\pm 0.004)$ , which corresponds to $\sim1.6^{\circ}$ tilt of the Ising axes of the three spins of the Kagome plane towards the plane [Fig. \ref{fig4}(c)]. This is consistent with the observed upshift of $\mathrm{B_c}$ because the projection of the field to these local spin axes becomes smaller and a larger field is necessary to flip the spin chain to stabilize (3:1).

    With this $\rm{\epsilon}$ value, the $\mathrm{B_c}=$ 14 T corresponds to $\rm{J_{eff}}=1.054\,K(\pm 0.171\,K)$ , which is close to the bulk value of $\rm{J_{eff}}=$ 1.01 K \cite{sato2007field}, albeit slightly bigger. Our previous studies \cite{zhou2012chemical} on bulk $\ch{Dy2Sn2O7}$, $\ch{Dy2Ti2O7}$, and $\ch{Dy2Ge2O7}$ show that $\rm{J_{eff}}$ decreases with a smaller unit cell volume \textcolor{black}{when both $\rm{J_{nn}}$ and $\rm{D_{nn}}$ increase. This is possible because $\rm{J_{nn}}$ is antiferromagnetic and increases faster than $\rm{D_{nn}}$ as $\rm{J_{nn}}$ is more sensitive to the bond length change than $\rm{D_{nn}}$.} The slight increase of $\rm{J_{eff}}$ in the DTO film could be attributed to the overall unit cell volume expansion (see End Matter) \textcolor{black}{following a similar trend}. The volume expansion is primarily caused by expansion in the (111) plane due to the lattice mismatch with the YSZ substrate without a full compensation from the compression along the [111] axis. Internally, the tetrahedron should experience a similar distortion, which is \textcolor{black}{expected to tilt the Dy-O bond between the Kagome-plane Dy site and the O site at the tetrahedral center toward the Kagome plane, as shown in Fig. S3(b) \cite{supplemental}. Since this Dy-O bond is where the Ising axis is \cite{tomasello2015single,ruminy2016crystal}, such a distortion is} consistent with a tilt of the Ising axes toward the Kagome plane as suggested by a positive $\rm{\epsilon}$. \textcolor{black}{This distortion corresponds to local symmetry-lowering around the central O site from $\rm{T_{d}}$ to $\rm{C_{3v}}$ as the epitaxial growth along [111] effectively lowers the pyrochlore symmetry from cubic to rhombohedral.} The phase boundary drawn by using Eq. \ref{eq2} is also more consistent with the experimental one, except a slight overestimation of $\mathrm{B_c}$ at small angles likely due to effects beyond the nearest-neighbor coupling \cite{anand2022investigation,borzi2016intermediate}.

    In summary, the proximitized transport and CTM results prove that the DTO thin film on YSZ still has the spin-ice ground state but with modified magnetic interactions likely due to a combination of volume expansion and geometric distortion. This is consistent with theoretical studies that suggest the persistence of the spin ice state within a substantial range of strain \cite{lu2024111}. \textcolor{black}{It will be an interest direction of future studies on whether the spin-ice manifold is fully retained. Locally, the six-fold degeneracy of a single tetrahedron should be fully retained because the three Dy sites on the Kagome plane are still equivalent. However, shape anisotropy alone, e.g., due to the thin-film geometry, may differentiate spin-ice network configurations that have a finite magnetization.} In general, our results show that both the CTM and the proximitized transport are able to probe thin layers of insulating frustrated magnets but from different perspectives. While the former detects the magnetization vector, the latter is sensitive to the fluctuations. Their consistency demonstrates that proximitized transport is an efficient probe, which could potentially be applicable even if the magnetic signal of the bulk of the film is beyond the CTM detection, since the response of the itinerant electrons only relies on the interfacial coupling with the local moments but not the sample volume. This approach could further enable thin film engineering of insulating quantum magnets even beyond classical spin ice \cite{xing2024anomalous,bovo2016layer,wen2021epitaxial}. It is also interesting to see whether the proximitized transport phenomenology would persist and/or modify when the conductive epilayer has spontaneous magnetic ordering \cite{wu2024electronic,kareev2025epitaxial}.

	Acknowledgment: This research was supported by the U.S. Department of Energy under Grant No. DE-SC0020254. Part of the work was done in the National High Magnetic Field Laboratory, which is supported by the National Science Foundation Cooperative Agreement No. DMR-1644779, and the State of Florida. The research at the University of Michigan was supported by the Department of Energy under Award No. DE-SC0020184 (magnetization torque measurement) to Kuan-Wen Chen, Guoxin Zheng, and Lu Li. Chengkun Xing received partial support from the Center for Material Processing at the University of Tennessee, Knoxville.
    \nocite{yang2017epitaxial}
    \nocite{apsrev42Control}

    \bibliography{reference}

\begin{thebibliography}{56}%
\makeatletter
\providecommand \@ifxundefined [1]{%
 \@ifx{#1\undefined}
}%
\providecommand \@ifnum [1]{%
 \ifnum #1\expandafter \@firstoftwo
 \else \expandafter \@secondoftwo
 \fi
}%
\providecommand \@ifx [1]{%
 \ifx #1\expandafter \@firstoftwo
 \else \expandafter \@secondoftwo
 \fi
}%
\providecommand \natexlab [1]{#1}%
\providecommand \enquote  [1]{``#1''}%
\providecommand \bibnamefont  [1]{#1}%
\providecommand \bibfnamefont [1]{#1}%
\providecommand \citenamefont [1]{#1}%
\providecommand \href@noop [0]{\@secondoftwo}%
\providecommand \href [0]{\begingroup \@sanitize@url \@href}%
\providecommand \@href[1]{\@@startlink{#1}\@@href}%
\providecommand \@@href[1]{\endgroup#1\@@endlink}%
\providecommand \@sanitize@url [0]{\catcode `\\12\catcode `\$12\catcode `\&12\catcode `\#12\catcode `\^12\catcode `\_12\catcode `\%12\relax}%
\providecommand \@@startlink[1]{}%
\providecommand \@@endlink[0]{}%
\providecommand \url  [0]{\begingroup\@sanitize@url \@url }%
\providecommand \@url [1]{\endgroup\@href {#1}{\urlprefix }}%
\providecommand \urlprefix  [0]{URL }%
\providecommand \Eprint [0]{\href }%
\providecommand \doibase [0]{https://doi.org/}%
\providecommand \selectlanguage [0]{\@gobble}%
\providecommand \bibinfo  [0]{\@secondoftwo}%
\providecommand \bibfield  [0]{\@secondoftwo}%
\providecommand \translation [1]{[#1]}%
\providecommand \BibitemOpen [0]{}%
\providecommand \bibitemStop [0]{}%
\providecommand \bibitemNoStop [0]{.\EOS\space}%
\providecommand \EOS [0]{\spacefactor3000\relax}%
\providecommand \BibitemShut  [1]{\csname bibitem#1\endcsname}%
\let\auto@bib@innerbib\@empty
\bibitem [{\citenamefont {Gardner}\ \emph {et~al.}(2010)\citenamefont {Gardner}, \citenamefont {Gingras},\ and\ \citenamefont {Greedan}}]{gardner2010magnetic}%
  \BibitemOpen
  \bibfield  {author} {\bibinfo {author} {\bibfnamefont {J.~S.}\ \bibnamefont {Gardner}}, \bibinfo {author} {\bibfnamefont {M.~J.}\ \bibnamefont {Gingras}},\ and\ \bibinfo {author} {\bibfnamefont {J.~E.}\ \bibnamefont {Greedan}},\ }\bibfield  {title} {\bibinfo {title} {Magnetic pyrochlore oxides},\ }\href@noop {} {\bibfield  {journal} {\bibinfo  {journal} {Reviews of Modern Physics}\ }\textbf {\bibinfo {volume} {82}},\ \bibinfo {pages} {53} (\bibinfo {year} {2010})}\BibitemShut {NoStop}%
\bibitem [{\citenamefont {Harris}\ \emph {et~al.}(1997)\citenamefont {Harris}, \citenamefont {Bramwell}, \citenamefont {McMorrow}, \citenamefont {Zeiske},\ and\ \citenamefont {Godfrey}}]{harris1997geometrical}%
  \BibitemOpen
  \bibfield  {author} {\bibinfo {author} {\bibfnamefont {M.~J.}\ \bibnamefont {Harris}}, \bibinfo {author} {\bibfnamefont {S.}~\bibnamefont {Bramwell}}, \bibinfo {author} {\bibfnamefont {D.}~\bibnamefont {McMorrow}}, \bibinfo {author} {\bibfnamefont {T.}~\bibnamefont {Zeiske}},\ and\ \bibinfo {author} {\bibfnamefont {K.}~\bibnamefont {Godfrey}},\ }\bibfield  {title} {\bibinfo {title} {Geometrical frustration in the ferromagnetic pyrochlore \ch{Ho2Ti2O7}},\ }\href@noop {} {\bibfield  {journal} {\bibinfo  {journal} {Physical Review Letters}\ }\textbf {\bibinfo {volume} {79}},\ \bibinfo {pages} {2554} (\bibinfo {year} {1997})}\BibitemShut {NoStop}%
\bibitem [{\citenamefont {Isakov}\ \emph {et~al.}(2005)\citenamefont {Isakov}, \citenamefont {Moessner},\ and\ \citenamefont {Sondhi}}]{isakov2005spin}%
  \BibitemOpen
  \bibfield  {author} {\bibinfo {author} {\bibfnamefont {S.~V.}\ \bibnamefont {Isakov}}, \bibinfo {author} {\bibfnamefont {R.}~\bibnamefont {Moessner}},\ and\ \bibinfo {author} {\bibfnamefont {S.~L.}\ \bibnamefont {Sondhi}},\ }\bibfield  {title} {\bibinfo {title} {Why spin ice obeys the ice rules},\ }\href@noop {} {\bibfield  {journal} {\bibinfo  {journal} {Physical Review Letters}\ }\textbf {\bibinfo {volume} {95}},\ \bibinfo {pages} {217201} (\bibinfo {year} {2005})}\BibitemShut {NoStop}%
\bibitem [{\citenamefont {Siddharthan}\ \emph {et~al.}(1999)\citenamefont {Siddharthan}, \citenamefont {Shastry}, \citenamefont {Ramirez}, \citenamefont {Hayashi}, \citenamefont {Cava},\ and\ \citenamefont {Rosenkranz}}]{siddharthan1999ising}%
  \BibitemOpen
  \bibfield  {author} {\bibinfo {author} {\bibfnamefont {R.}~\bibnamefont {Siddharthan}}, \bibinfo {author} {\bibfnamefont {B.}~\bibnamefont {Shastry}}, \bibinfo {author} {\bibfnamefont {A.}~\bibnamefont {Ramirez}}, \bibinfo {author} {\bibfnamefont {A.}~\bibnamefont {Hayashi}}, \bibinfo {author} {\bibfnamefont {R.}~\bibnamefont {Cava}},\ and\ \bibinfo {author} {\bibfnamefont {S.}~\bibnamefont {Rosenkranz}},\ }\bibfield  {title} {\bibinfo {title} {Ising pyrochlore magnets: Low-temperature properties,“ice rules,” and beyond},\ }\href@noop {} {\bibfield  {journal} {\bibinfo  {journal} {Physical Review Letters}\ }\textbf {\bibinfo {volume} {83}},\ \bibinfo {pages} {1854} (\bibinfo {year} {1999})}\BibitemShut {NoStop}%
\bibitem [{\citenamefont {Bramwell}\ and\ \citenamefont {Harris}(1998)}]{bramwell1998frustration}%
  \BibitemOpen
  \bibfield  {author} {\bibinfo {author} {\bibfnamefont {S.}~\bibnamefont {Bramwell}}\ and\ \bibinfo {author} {\bibfnamefont {M.}~\bibnamefont {Harris}},\ }\bibfield  {title} {\bibinfo {title} {Frustration in ising-type spin models on the pyrochlore lattice},\ }\href@noop {} {\bibfield  {journal} {\bibinfo  {journal} {Journal of Physics: Condensed Matter}\ }\textbf {\bibinfo {volume} {10}},\ \bibinfo {pages} {L215} (\bibinfo {year} {1998})}\BibitemShut {NoStop}%
\bibitem [{\citenamefont {Bramwell}\ and\ \citenamefont {Harris}(2020)}]{bramwell2020history}%
  \BibitemOpen
  \bibfield  {author} {\bibinfo {author} {\bibfnamefont {S.~T.}\ \bibnamefont {Bramwell}}\ and\ \bibinfo {author} {\bibfnamefont {M.~J.}\ \bibnamefont {Harris}},\ }\bibfield  {title} {\bibinfo {title} {The history of spin ice},\ }\href@noop {} {\bibfield  {journal} {\bibinfo  {journal} {Journal of Physics: Condensed Matter}\ }\textbf {\bibinfo {volume} {32}},\ \bibinfo {pages} {374010} (\bibinfo {year} {2020})}\BibitemShut {NoStop}%
\bibitem [{\citenamefont {Snyder}\ \emph {et~al.}(2001)\citenamefont {Snyder}, \citenamefont {Slusky}, \citenamefont {Cava},\ and\ \citenamefont {Schiffer}}]{snyder2001spin}%
  \BibitemOpen
  \bibfield  {author} {\bibinfo {author} {\bibfnamefont {J.}~\bibnamefont {Snyder}}, \bibinfo {author} {\bibfnamefont {J.}~\bibnamefont {Slusky}}, \bibinfo {author} {\bibfnamefont {R.}~\bibnamefont {Cava}},\ and\ \bibinfo {author} {\bibfnamefont {P.}~\bibnamefont {Schiffer}},\ }\bibfield  {title} {\bibinfo {title} {How ‘spin ice’freezes},\ }\href@noop {} {\bibfield  {journal} {\bibinfo  {journal} {Nature}\ }\textbf {\bibinfo {volume} {413}},\ \bibinfo {pages} {48} (\bibinfo {year} {2001})}\BibitemShut {NoStop}%
\bibitem [{\citenamefont {Pauling}(1935)}]{pauling1935structure}%
  \BibitemOpen
  \bibfield  {author} {\bibinfo {author} {\bibfnamefont {L.}~\bibnamefont {Pauling}},\ }\bibfield  {title} {\bibinfo {title} {The structure and entropy of ice and of other crystals with some randomness of atomic arrangement},\ }\href@noop {} {\bibfield  {journal} {\bibinfo  {journal} {Journal of the American Chemical Society}\ }\textbf {\bibinfo {volume} {57}},\ \bibinfo {pages} {2680} (\bibinfo {year} {1935})}\BibitemShut {NoStop}%
\bibitem [{\citenamefont {Castelnovo}\ \emph {et~al.}(2008)\citenamefont {Castelnovo}, \citenamefont {Moessner},\ and\ \citenamefont {Sondhi}}]{castelnovo2008magnetic}%
  \BibitemOpen
  \bibfield  {author} {\bibinfo {author} {\bibfnamefont {C.}~\bibnamefont {Castelnovo}}, \bibinfo {author} {\bibfnamefont {R.}~\bibnamefont {Moessner}},\ and\ \bibinfo {author} {\bibfnamefont {S.~L.}\ \bibnamefont {Sondhi}},\ }\bibfield  {title} {\bibinfo {title} {Magnetic monopoles in spin ice},\ }\href@noop {} {\bibfield  {journal} {\bibinfo  {journal} {Nature}\ }\textbf {\bibinfo {volume} {451}},\ \bibinfo {pages} {42} (\bibinfo {year} {2008})}\BibitemShut {NoStop}%
\bibitem [{\citenamefont {Fennell}\ \emph {et~al.}(2009)\citenamefont {Fennell}, \citenamefont {Deen}, \citenamefont {Wildes}, \citenamefont {Schmalzl}, \citenamefont {Prabhakaran}, \citenamefont {Boothroyd}, \citenamefont {Aldus}, \citenamefont {McMorrow},\ and\ \citenamefont {Bramwell}}]{fennell2009magnetic}%
  \BibitemOpen
  \bibfield  {author} {\bibinfo {author} {\bibfnamefont {T.}~\bibnamefont {Fennell}}, \bibinfo {author} {\bibfnamefont {P.}~\bibnamefont {Deen}}, \bibinfo {author} {\bibfnamefont {A.}~\bibnamefont {Wildes}}, \bibinfo {author} {\bibfnamefont {K.}~\bibnamefont {Schmalzl}}, \bibinfo {author} {\bibfnamefont {D.}~\bibnamefont {Prabhakaran}}, \bibinfo {author} {\bibfnamefont {A.}~\bibnamefont {Boothroyd}}, \bibinfo {author} {\bibfnamefont {R.}~\bibnamefont {Aldus}}, \bibinfo {author} {\bibfnamefont {D.}~\bibnamefont {McMorrow}},\ and\ \bibinfo {author} {\bibfnamefont {S.}~\bibnamefont {Bramwell}},\ }\bibfield  {title} {\bibinfo {title} {Magnetic coulomb phase in the spin ice \ch{Ho2Ti2O7}},\ }\href@noop {} {\bibfield  {journal} {\bibinfo  {journal} {Science}\ }\textbf {\bibinfo {volume} {326}},\ \bibinfo {pages} {415} (\bibinfo {year} {2009})}\BibitemShut {NoStop}%
\bibitem [{\citenamefont {Jaubert}\ and\ \citenamefont {Holdsworth}(2011)}]{jaubert2011magnetic}%
  \BibitemOpen
  \bibfield  {author} {\bibinfo {author} {\bibfnamefont {L.~D.}\ \bibnamefont {Jaubert}}\ and\ \bibinfo {author} {\bibfnamefont {P.~C.}\ \bibnamefont {Holdsworth}},\ }\bibfield  {title} {\bibinfo {title} {Magnetic monopole dynamics in spin ice},\ }\href@noop {} {\bibfield  {journal} {\bibinfo  {journal} {Journal of Physics: Condensed Matter}\ }\textbf {\bibinfo {volume} {23}},\ \bibinfo {pages} {164222} (\bibinfo {year} {2011})}\BibitemShut {NoStop}%
\bibitem [{\citenamefont {Jaubert}\ and\ \citenamefont {Holdsworth}(2009)}]{jaubert2009signature}%
  \BibitemOpen
  \bibfield  {author} {\bibinfo {author} {\bibfnamefont {L.~D.}\ \bibnamefont {Jaubert}}\ and\ \bibinfo {author} {\bibfnamefont {P.~C.}\ \bibnamefont {Holdsworth}},\ }\bibfield  {title} {\bibinfo {title} {Signature of magnetic monopole and dirac string dynamics in spin ice},\ }\href@noop {} {\bibfield  {journal} {\bibinfo  {journal} {Nature Physics}\ }\textbf {\bibinfo {volume} {5}},\ \bibinfo {pages} {258} (\bibinfo {year} {2009})}\BibitemShut {NoStop}%
\bibitem [{\citenamefont {Morris}\ \emph {et~al.}(2009)\citenamefont {Morris}, \citenamefont {Tennant}, \citenamefont {Grigera}, \citenamefont {Klemke}, \citenamefont {Castelnovo}, \citenamefont {Moessner}, \citenamefont {Czternasty}, \citenamefont {Meissner}, \citenamefont {Rule}, \citenamefont {Hoffmann} \emph {et~al.}}]{morris2009dirac}%
  \BibitemOpen
  \bibfield  {author} {\bibinfo {author} {\bibfnamefont {D.~J.~P.}\ \bibnamefont {Morris}}, \bibinfo {author} {\bibfnamefont {D.}~\bibnamefont {Tennant}}, \bibinfo {author} {\bibfnamefont {S.}~\bibnamefont {Grigera}}, \bibinfo {author} {\bibfnamefont {B.}~\bibnamefont {Klemke}}, \bibinfo {author} {\bibfnamefont {C.}~\bibnamefont {Castelnovo}}, \bibinfo {author} {\bibfnamefont {R.}~\bibnamefont {Moessner}}, \bibinfo {author} {\bibfnamefont {C.}~\bibnamefont {Czternasty}}, \bibinfo {author} {\bibfnamefont {M.}~\bibnamefont {Meissner}}, \bibinfo {author} {\bibfnamefont {K.}~\bibnamefont {Rule}}, \bibinfo {author} {\bibfnamefont {J.-U.}\ \bibnamefont {Hoffmann}}, \emph {et~al.},\ }\bibfield  {title} {\bibinfo {title} {Dirac strings and magnetic monopoles in the spin ice \ch{Dy2Ti2O7}},\ }\href@noop {} {\bibfield  {journal} {\bibinfo  {journal} {Science}\ }\textbf {\bibinfo {volume} {326}},\ \bibinfo {pages} {411} (\bibinfo {year} {2009})}\BibitemShut {NoStop}%
\bibitem [{\citenamefont {Khomskii}(2012)}]{khomskii2012electric}%
  \BibitemOpen
  \bibfield  {author} {\bibinfo {author} {\bibfnamefont {D.}~\bibnamefont {Khomskii}},\ }\bibfield  {title} {\bibinfo {title} {Electric dipoles on magnetic monopoles in spin ice},\ }\href@noop {} {\bibfield  {journal} {\bibinfo  {journal} {Nature communications}\ }\textbf {\bibinfo {volume} {3}},\ \bibinfo {pages} {904} (\bibinfo {year} {2012})}\BibitemShut {NoStop}%
\bibitem [{\citenamefont {Tomasello}\ \emph {et~al.}(2015)\citenamefont {Tomasello}, \citenamefont {Castelnovo}, \citenamefont {Moessner},\ and\ \citenamefont {Quintanilla}}]{tomasello2015single}%
  \BibitemOpen
  \bibfield  {author} {\bibinfo {author} {\bibfnamefont {B.}~\bibnamefont {Tomasello}}, \bibinfo {author} {\bibfnamefont {C.}~\bibnamefont {Castelnovo}}, \bibinfo {author} {\bibfnamefont {R.}~\bibnamefont {Moessner}},\ and\ \bibinfo {author} {\bibfnamefont {J.}~\bibnamefont {Quintanilla}},\ }\bibfield  {title} {\bibinfo {title} {Single-ion anisotropy and magnetic field response in the spin-ice materials \ch{Ho2Ti2O7} and \ch{Dy2Ti2O7}},\ }\href@noop {} {\bibfield  {journal} {\bibinfo  {journal} {Physical Review B}\ }\textbf {\bibinfo {volume} {92}},\ \bibinfo {pages} {155120} (\bibinfo {year} {2015})}\BibitemShut {NoStop}%
\bibitem [{\citenamefont {Jana}\ \emph {et~al.}(2002)\citenamefont {Jana}, \citenamefont {Sengupta},\ and\ \citenamefont {Ghosh}}]{jana2002estimation}%
  \BibitemOpen
  \bibfield  {author} {\bibinfo {author} {\bibfnamefont {Y.}~\bibnamefont {Jana}}, \bibinfo {author} {\bibfnamefont {A.}~\bibnamefont {Sengupta}},\ and\ \bibinfo {author} {\bibfnamefont {D.}~\bibnamefont {Ghosh}},\ }\bibfield  {title} {\bibinfo {title} {Estimation of single ion anisotropy in pyrochlore \ch{Dy2Ti2O7}, a geometrically frustrated system, using crystal field theory},\ }\href@noop {} {\bibfield  {journal} {\bibinfo  {journal} {Journal of magnetism and magnetic materials}\ }\textbf {\bibinfo {volume} {248}},\ \bibinfo {pages} {7} (\bibinfo {year} {2002})}\BibitemShut {NoStop}%
\bibitem [{\citenamefont {Bramwell}\ \emph {et~al.}(2001)\citenamefont {Bramwell}, \citenamefont {Harris}, \citenamefont {Den~Hertog}, \citenamefont {Gingras}, \citenamefont {Gardner}, \citenamefont {McMorrow}, \citenamefont {Wildes}, \citenamefont {Cornelius}, \citenamefont {Champion}, \citenamefont {Melko} \emph {et~al.}}]{bramwell2001spin}%
  \BibitemOpen
  \bibfield  {author} {\bibinfo {author} {\bibfnamefont {S.}~\bibnamefont {Bramwell}}, \bibinfo {author} {\bibfnamefont {M.}~\bibnamefont {Harris}}, \bibinfo {author} {\bibfnamefont {B.}~\bibnamefont {Den~Hertog}}, \bibinfo {author} {\bibfnamefont {M.}~\bibnamefont {Gingras}}, \bibinfo {author} {\bibfnamefont {J.}~\bibnamefont {Gardner}}, \bibinfo {author} {\bibfnamefont {D.}~\bibnamefont {McMorrow}}, \bibinfo {author} {\bibfnamefont {A.}~\bibnamefont {Wildes}}, \bibinfo {author} {\bibfnamefont {A.}~\bibnamefont {Cornelius}}, \bibinfo {author} {\bibfnamefont {J.}~\bibnamefont {Champion}}, \bibinfo {author} {\bibfnamefont {R.}~\bibnamefont {Melko}}, \emph {et~al.},\ }\bibfield  {title} {\bibinfo {title} {Spin correlations in \ch{Ho2Ti2O7}: a dipolar spin ice system},\ }\href@noop {} {\bibfield  {journal} {\bibinfo  {journal} {Physical Review Letters}\ }\textbf {\bibinfo {volume} {87}},\ \bibinfo {pages} {047205} (\bibinfo {year} {2001})}\BibitemShut {NoStop}%
\bibitem [{\citenamefont {Ruff}\ \emph {et~al.}(2005)\citenamefont {Ruff}, \citenamefont {Melko},\ and\ \citenamefont {Gingras}}]{ruff2005finite}%
  \BibitemOpen
  \bibfield  {author} {\bibinfo {author} {\bibfnamefont {J.~P.}\ \bibnamefont {Ruff}}, \bibinfo {author} {\bibfnamefont {R.~G.}\ \bibnamefont {Melko}},\ and\ \bibinfo {author} {\bibfnamefont {M.~J.}\ \bibnamefont {Gingras}},\ }\bibfield  {title} {\bibinfo {title} {Finite-temperature transitions in dipolar spin ice in a large magnetic field},\ }\href@noop {} {\bibfield  {journal} {\bibinfo  {journal} {Physical Review Letters}\ }\textbf {\bibinfo {volume} {95}},\ \bibinfo {pages} {097202} (\bibinfo {year} {2005})}\BibitemShut {NoStop}%
\bibitem [{\citenamefont {den Hertog}\ and\ \citenamefont {Gingras}(2000)}]{den2000dipolar}%
  \BibitemOpen
  \bibfield  {author} {\bibinfo {author} {\bibfnamefont {B.~C.}\ \bibnamefont {den Hertog}}\ and\ \bibinfo {author} {\bibfnamefont {M.~J.}\ \bibnamefont {Gingras}},\ }\bibfield  {title} {\bibinfo {title} {Dipolar interactions and origin of spin ice in ising pyrochlore magnets},\ }\href@noop {} {\bibfield  {journal} {\bibinfo  {journal} {Physical Review Letters}\ }\textbf {\bibinfo {volume} {84}},\ \bibinfo {pages} {3430} (\bibinfo {year} {2000})}\BibitemShut {NoStop}%
\bibitem [{\citenamefont {Bramwell}\ and\ \citenamefont {Gingras}(2001)}]{bramwell2001spinscience}%
  \BibitemOpen
  \bibfield  {author} {\bibinfo {author} {\bibfnamefont {S.~T.}\ \bibnamefont {Bramwell}}\ and\ \bibinfo {author} {\bibfnamefont {M.~J.}\ \bibnamefont {Gingras}},\ }\bibfield  {title} {\bibinfo {title} {Spin ice state in frustrated magnetic pyrochlore materials},\ }\href@noop {} {\bibfield  {journal} {\bibinfo  {journal} {Science}\ }\textbf {\bibinfo {volume} {294}},\ \bibinfo {pages} {1495} (\bibinfo {year} {2001})}\BibitemShut {NoStop}%
\bibitem [{\citenamefont {Ramirez}\ \emph {et~al.}(1999)\citenamefont {Ramirez}, \citenamefont {Hayashi}, \citenamefont {Cava}, \citenamefont {Siddharthan},\ and\ \citenamefont {Shastry}}]{ramirez1999zero}%
  \BibitemOpen
  \bibfield  {author} {\bibinfo {author} {\bibfnamefont {A.~P.}\ \bibnamefont {Ramirez}}, \bibinfo {author} {\bibfnamefont {A.}~\bibnamefont {Hayashi}}, \bibinfo {author} {\bibfnamefont {R.~J.}\ \bibnamefont {Cava}}, \bibinfo {author} {\bibfnamefont {R.}~\bibnamefont {Siddharthan}},\ and\ \bibinfo {author} {\bibfnamefont {B.}~\bibnamefont {Shastry}},\ }\bibfield  {title} {\bibinfo {title} {Zero-point entropy in ‘spin ice’},\ }\href@noop {} {\bibfield  {journal} {\bibinfo  {journal} {Nature}\ }\textbf {\bibinfo {volume} {399}},\ \bibinfo {pages} {333} (\bibinfo {year} {1999})}\BibitemShut {NoStop}%
\bibitem [{\citenamefont {Sakakibara}\ \emph {et~al.}(2003)\citenamefont {Sakakibara}, \citenamefont {Tayama}, \citenamefont {Hiroi}, \citenamefont {Matsuhira},\ and\ \citenamefont {Takagi}}]{sakakibara2003observation}%
  \BibitemOpen
  \bibfield  {author} {\bibinfo {author} {\bibfnamefont {T.}~\bibnamefont {Sakakibara}}, \bibinfo {author} {\bibfnamefont {T.}~\bibnamefont {Tayama}}, \bibinfo {author} {\bibfnamefont {Z.}~\bibnamefont {Hiroi}}, \bibinfo {author} {\bibfnamefont {K.}~\bibnamefont {Matsuhira}},\ and\ \bibinfo {author} {\bibfnamefont {S.}~\bibnamefont {Takagi}},\ }\bibfield  {title} {\bibinfo {title} {Observation of a liquid-gas-type transition in the pyrochlore spin ice compound \ch{Dy2Ti2O7} in a magnetic field},\ }\href@noop {} {\bibfield  {journal} {\bibinfo  {journal} {Physical Review Letters}\ }\textbf {\bibinfo {volume} {90}},\ \bibinfo {pages} {207205} (\bibinfo {year} {2003})}\BibitemShut {NoStop}%
\bibitem [{\citenamefont {Fukazawa}\ \emph {et~al.}(2002)\citenamefont {Fukazawa}, \citenamefont {Melko}, \citenamefont {Higashinaka}, \citenamefont {Maeno},\ and\ \citenamefont {Gingras}}]{fukazawa2002magnetic}%
  \BibitemOpen
  \bibfield  {author} {\bibinfo {author} {\bibfnamefont {H.}~\bibnamefont {Fukazawa}}, \bibinfo {author} {\bibfnamefont {R.}~\bibnamefont {Melko}}, \bibinfo {author} {\bibfnamefont {R.}~\bibnamefont {Higashinaka}}, \bibinfo {author} {\bibfnamefont {Y.}~\bibnamefont {Maeno}},\ and\ \bibinfo {author} {\bibfnamefont {M.}~\bibnamefont {Gingras}},\ }\bibfield  {title} {\bibinfo {title} {Magnetic anisotropy of the spin-ice compound dy 2 ti 2 o 7},\ }\href@noop {} {\bibfield  {journal} {\bibinfo  {journal} {Physical Review B}\ }\textbf {\bibinfo {volume} {65}},\ \bibinfo {pages} {054410} (\bibinfo {year} {2002})}\BibitemShut {NoStop}%
\bibitem [{\citenamefont {Petrenko}\ \emph {et~al.}(2003)\citenamefont {Petrenko}, \citenamefont {Lees},\ and\ \citenamefont {Balakrishnan}}]{petrenko2003magnetization}%
  \BibitemOpen
  \bibfield  {author} {\bibinfo {author} {\bibfnamefont {O.}~\bibnamefont {Petrenko}}, \bibinfo {author} {\bibfnamefont {M.}~\bibnamefont {Lees}},\ and\ \bibinfo {author} {\bibfnamefont {G.}~\bibnamefont {Balakrishnan}},\ }\bibfield  {title} {\bibinfo {title} {Magnetization process in the spin-ice compound \ch{Ho2Ti2O7}},\ }\href@noop {} {\bibfield  {journal} {\bibinfo  {journal} {Physical Review B}\ }\textbf {\bibinfo {volume} {68}},\ \bibinfo {pages} {012406} (\bibinfo {year} {2003})}\BibitemShut {NoStop}%
\bibitem [{\citenamefont {Hiroi}\ \emph {et~al.}(2003)\citenamefont {Hiroi}, \citenamefont {Matsuhira}, \citenamefont {Takagi}, \citenamefont {Tayama},\ and\ \citenamefont {Sakakibara}}]{hiroi2003specific}%
  \BibitemOpen
  \bibfield  {author} {\bibinfo {author} {\bibfnamefont {Z.}~\bibnamefont {Hiroi}}, \bibinfo {author} {\bibfnamefont {K.}~\bibnamefont {Matsuhira}}, \bibinfo {author} {\bibfnamefont {S.}~\bibnamefont {Takagi}}, \bibinfo {author} {\bibfnamefont {T.}~\bibnamefont {Tayama}},\ and\ \bibinfo {author} {\bibfnamefont {T.}~\bibnamefont {Sakakibara}},\ }\bibfield  {title} {\bibinfo {title} {Specific heat of kagome ice in the pyrochlore oxide \ch{Dy2Ti2O7}},\ }\href@noop {} {\bibfield  {journal} {\bibinfo  {journal} {Journal of the Physical Society of Japan}\ }\textbf {\bibinfo {volume} {72}},\ \bibinfo {pages} {411} (\bibinfo {year} {2003})}\BibitemShut {NoStop}%
\bibitem [{\citenamefont {Higashinaka}\ \emph {et~al.}(2004)\citenamefont {Higashinaka}, \citenamefont {Fukazawa}, \citenamefont {Deguchi},\ and\ \citenamefont {Maeno}}]{higashinaka2004low}%
  \BibitemOpen
  \bibfield  {author} {\bibinfo {author} {\bibfnamefont {R.}~\bibnamefont {Higashinaka}}, \bibinfo {author} {\bibfnamefont {H.}~\bibnamefont {Fukazawa}}, \bibinfo {author} {\bibfnamefont {K.}~\bibnamefont {Deguchi}},\ and\ \bibinfo {author} {\bibfnamefont {Y.}~\bibnamefont {Maeno}},\ }\bibfield  {title} {\bibinfo {title} {Low temperature specific heat of \ch{Dy2Ti2O7} in the kagome ice state},\ }\href@noop {} {\bibfield  {journal} {\bibinfo  {journal} {Journal of the Physical Society of Japan}\ }\textbf {\bibinfo {volume} {73}},\ \bibinfo {pages} {2845} (\bibinfo {year} {2004})}\BibitemShut {NoStop}%
\bibitem [{\citenamefont {Aoki}\ \emph {et~al.}(2004)\citenamefont {Aoki}, \citenamefont {Sakakibara}, \citenamefont {Matsuhira},\ and\ \citenamefont {Hiroi}}]{aoki2004magnetocaloric}%
  \BibitemOpen
  \bibfield  {author} {\bibinfo {author} {\bibfnamefont {H.}~\bibnamefont {Aoki}}, \bibinfo {author} {\bibfnamefont {T.}~\bibnamefont {Sakakibara}}, \bibinfo {author} {\bibfnamefont {K.}~\bibnamefont {Matsuhira}},\ and\ \bibinfo {author} {\bibfnamefont {Z.}~\bibnamefont {Hiroi}},\ }\bibfield  {title} {\bibinfo {title} {Magnetocaloric effect study on the pyrochlore spin ice compound \ch{Dy2Ti2O7} in a [111] magnetic field},\ }\href@noop {} {\bibfield  {journal} {\bibinfo  {journal} {Journal of the Physical Society of Japan}\ }\textbf {\bibinfo {volume} {73}},\ \bibinfo {pages} {2851} (\bibinfo {year} {2004})}\BibitemShut {NoStop}%
\bibitem [{\citenamefont {Grigera}\ \emph {et~al.}(2015)\citenamefont {Grigera}, \citenamefont {Borzi}, \citenamefont {Slobinsky}, \citenamefont {Gibbs}, \citenamefont {Higashinaka}, \citenamefont {Maeno},\ and\ \citenamefont {Grigera}}]{grigera2015intermediate}%
  \BibitemOpen
  \bibfield  {author} {\bibinfo {author} {\bibfnamefont {S.~A.}\ \bibnamefont {Grigera}}, \bibinfo {author} {\bibfnamefont {R.~A.}\ \bibnamefont {Borzi}}, \bibinfo {author} {\bibfnamefont {D.~G.}\ \bibnamefont {Slobinsky}}, \bibinfo {author} {\bibfnamefont {A.}~\bibnamefont {Gibbs}}, \bibinfo {author} {\bibfnamefont {R.}~\bibnamefont {Higashinaka}}, \bibinfo {author} {\bibfnamefont {Y.}~\bibnamefont {Maeno}},\ and\ \bibinfo {author} {\bibfnamefont {T.~S.}\ \bibnamefont {Grigera}},\ }\bibfield  {title} {\bibinfo {title} {An intermediate state between the kagome-ice and the fully polarized state in \ch{Dy2Ti2O7}},\ }\href@noop {} {\bibfield  {journal} {\bibinfo  {journal} {Papers in physics}\ }\textbf {\bibinfo {volume} {7}},\ \bibinfo {pages} {0} (\bibinfo {year} {2015})}\BibitemShut {NoStop}%
\bibitem [{\citenamefont {Bergmann}(1984)}]{bergmann1984weak}%
  \BibitemOpen
  \bibfield  {author} {\bibinfo {author} {\bibfnamefont {G.}~\bibnamefont {Bergmann}},\ }\bibfield  {title} {\bibinfo {title} {Weak localization in thin films: a time-of-flight experiment with conduction electrons},\ }\href@noop {} {\bibfield  {journal} {\bibinfo  {journal} {Physics Reports}\ }\textbf {\bibinfo {volume} {107}},\ \bibinfo {pages} {1} (\bibinfo {year} {1984})}\BibitemShut {NoStop}%
\bibitem [{\citenamefont {Pandey}\ \emph {et~al.}(2022)\citenamefont {Pandey}, \citenamefont {Zhang}, \citenamefont {Yang}, \citenamefont {May}, \citenamefont {Sanchez}, \citenamefont {Liu}, \citenamefont {Chu}, \citenamefont {Kim}, \citenamefont {Ryan}, \citenamefont {Zhou} \emph {et~al.}}]{pandey2022controllable}%
  \BibitemOpen
  \bibfield  {author} {\bibinfo {author} {\bibfnamefont {S.}~\bibnamefont {Pandey}}, \bibinfo {author} {\bibfnamefont {H.}~\bibnamefont {Zhang}}, \bibinfo {author} {\bibfnamefont {J.}~\bibnamefont {Yang}}, \bibinfo {author} {\bibfnamefont {A.~F.}\ \bibnamefont {May}}, \bibinfo {author} {\bibfnamefont {J.~J.}\ \bibnamefont {Sanchez}}, \bibinfo {author} {\bibfnamefont {Z.}~\bibnamefont {Liu}}, \bibinfo {author} {\bibfnamefont {J.-H.}\ \bibnamefont {Chu}}, \bibinfo {author} {\bibfnamefont {J.-W.}\ \bibnamefont {Kim}}, \bibinfo {author} {\bibfnamefont {P.~J.}\ \bibnamefont {Ryan}}, \bibinfo {author} {\bibfnamefont {H.}~\bibnamefont {Zhou}}, \emph {et~al.},\ }\bibfield  {title} {\bibinfo {title} {Controllable emergent spatial spin modulation in \ch{Sr2IrO4} by in situ shear strain},\ }\href@noop {} {\bibfield  {journal} {\bibinfo  {journal} {Physical Review Letters}\ }\textbf {\bibinfo {volume} {129}},\ \bibinfo {pages} {027203} (\bibinfo {year} {2022})}\BibitemShut {NoStop}%
\bibitem [{\citenamefont {Barry}\ \emph {et~al.}(2019)\citenamefont {Barry}, \citenamefont {Zhang}, \citenamefont {Anand}, \citenamefont {Xin}, \citenamefont {Vailionis}, \citenamefont {Neu}, \citenamefont {Heikes}, \citenamefont {Cochran}, \citenamefont {Zhou}, \citenamefont {Qiu} \emph {et~al.}}]{barry2019modification}%
  \BibitemOpen
  \bibfield  {author} {\bibinfo {author} {\bibfnamefont {K.}~\bibnamefont {Barry}}, \bibinfo {author} {\bibfnamefont {B.}~\bibnamefont {Zhang}}, \bibinfo {author} {\bibfnamefont {N.}~\bibnamefont {Anand}}, \bibinfo {author} {\bibfnamefont {Y.}~\bibnamefont {Xin}}, \bibinfo {author} {\bibfnamefont {A.}~\bibnamefont {Vailionis}}, \bibinfo {author} {\bibfnamefont {J.}~\bibnamefont {Neu}}, \bibinfo {author} {\bibfnamefont {C.}~\bibnamefont {Heikes}}, \bibinfo {author} {\bibfnamefont {C.}~\bibnamefont {Cochran}}, \bibinfo {author} {\bibfnamefont {H.}~\bibnamefont {Zhou}}, \bibinfo {author} {\bibfnamefont {Y.}~\bibnamefont {Qiu}}, \emph {et~al.},\ }\bibfield  {title} {\bibinfo {title} {Modification of spin-ice physics in \ch{Ho2Ti2O7} thin films},\ }\href@noop {} {\bibfield  {journal} {\bibinfo  {journal} {Physical Review Materials}\ }\textbf {\bibinfo {volume} {3}},\ \bibinfo {pages} {084412} (\bibinfo {year} {2019})}\BibitemShut {NoStop}%
\bibitem [{\citenamefont {Leusink}\ \emph {et~al.}(2014)\citenamefont {Leusink}, \citenamefont {Coneri}, \citenamefont {Hoek}, \citenamefont {Turner}, \citenamefont {Van~Tendeloo},\ and\ \citenamefont {Idrissi}}]{leusink2014thin}%
  \BibitemOpen
  \bibfield  {author} {\bibinfo {author} {\bibfnamefont {D.}~\bibnamefont {Leusink}}, \bibinfo {author} {\bibfnamefont {F.}~\bibnamefont {Coneri}}, \bibinfo {author} {\bibfnamefont {M.}~\bibnamefont {Hoek}}, \bibinfo {author} {\bibfnamefont {S.}~\bibnamefont {Turner}}, \bibinfo {author} {\bibfnamefont {G.}~\bibnamefont {Van~Tendeloo}},\ and\ \bibinfo {author} {\bibfnamefont {H.}~\bibnamefont {Idrissi}},\ }\bibfield  {title} {\bibinfo {title} {Thin films of the spin ice compound \ch{Ho2Ti2O7}},\ }\href@noop {} {\bibfield  {journal} {\bibinfo  {journal} {APL Materials}\ }\textbf {\bibinfo {volume} {2}} (\bibinfo {year} {2014})}\BibitemShut {NoStop}%
\bibitem [{\citenamefont {Bovo}\ \emph {et~al.}(2014)\citenamefont {Bovo}, \citenamefont {Moya}, \citenamefont {Prabhakaran}, \citenamefont {Soh}, \citenamefont {Boothroyd}, \citenamefont {Mathur}, \citenamefont {Aeppli},\ and\ \citenamefont {Bramwell}}]{bovo2014restoration}%
  \BibitemOpen
  \bibfield  {author} {\bibinfo {author} {\bibfnamefont {L.}~\bibnamefont {Bovo}}, \bibinfo {author} {\bibfnamefont {X.}~\bibnamefont {Moya}}, \bibinfo {author} {\bibfnamefont {D.}~\bibnamefont {Prabhakaran}}, \bibinfo {author} {\bibfnamefont {Y.-A.}\ \bibnamefont {Soh}}, \bibinfo {author} {\bibfnamefont {A.}~\bibnamefont {Boothroyd}}, \bibinfo {author} {\bibfnamefont {N.}~\bibnamefont {Mathur}}, \bibinfo {author} {\bibfnamefont {G.}~\bibnamefont {Aeppli}},\ and\ \bibinfo {author} {\bibfnamefont {S.}~\bibnamefont {Bramwell}},\ }\bibfield  {title} {\bibinfo {title} {Restoration of the third law in spin ice thin films},\ }\href@noop {} {\bibfield  {journal} {\bibinfo  {journal} {Nature Communications}\ }\textbf {\bibinfo {volume} {5}},\ \bibinfo {pages} {3439} (\bibinfo {year} {2014})}\BibitemShut {NoStop}%
\bibitem [{\citenamefont {Bovo}\ \emph {et~al.}(2019)\citenamefont {Bovo}, \citenamefont {Rouleau}, \citenamefont {Prabhakaran},\ and\ \citenamefont {Bramwell}}]{bovo2019phase}%
  \BibitemOpen
  \bibfield  {author} {\bibinfo {author} {\bibfnamefont {L.}~\bibnamefont {Bovo}}, \bibinfo {author} {\bibfnamefont {C.}~\bibnamefont {Rouleau}}, \bibinfo {author} {\bibfnamefont {D.}~\bibnamefont {Prabhakaran}},\ and\ \bibinfo {author} {\bibfnamefont {S.}~\bibnamefont {Bramwell}},\ }\bibfield  {title} {\bibinfo {title} {Phase transitions in few-monolayer spin ice films},\ }\href@noop {} {\bibfield  {journal} {\bibinfo  {journal} {Nature Communications}\ }\textbf {\bibinfo {volume} {10}},\ \bibinfo {pages} {1219} (\bibinfo {year} {2019})}\BibitemShut {NoStop}%
\bibitem [{\citenamefont {Wen}\ \emph {et~al.}(2021)\citenamefont {Wen}, \citenamefont {Wu}, \citenamefont {Liu}, \citenamefont {Terilli}, \citenamefont {Kareev},\ and\ \citenamefont {Chakhalian}}]{wen2021epitaxial}%
  \BibitemOpen
  \bibfield  {author} {\bibinfo {author} {\bibfnamefont {F.}~\bibnamefont {Wen}}, \bibinfo {author} {\bibfnamefont {T.-C.}\ \bibnamefont {Wu}}, \bibinfo {author} {\bibfnamefont {X.}~\bibnamefont {Liu}}, \bibinfo {author} {\bibfnamefont {M.}~\bibnamefont {Terilli}}, \bibinfo {author} {\bibfnamefont {M.}~\bibnamefont {Kareev}},\ and\ \bibinfo {author} {\bibfnamefont {J.}~\bibnamefont {Chakhalian}},\ }\bibfield  {title} {\bibinfo {title} {Epitaxial stabilization of (111)-oriented frustrated quantum pyrochlore thin films},\ }\href@noop {} {\bibfield  {journal} {\bibinfo  {journal} {Journal of Applied Physics}\ }\textbf {\bibinfo {volume} {129}},\ \bibinfo {pages} {025302} (\bibinfo {year} {2021})}\BibitemShut {NoStop}%
\bibitem [{\citenamefont {Miao}\ \emph {et~al.}(2020)\citenamefont {Miao}, \citenamefont {Lee}, \citenamefont {Mei}, \citenamefont {Lawler},\ and\ \citenamefont {Shen}}]{miao2020two}%
  \BibitemOpen
  \bibfield  {author} {\bibinfo {author} {\bibfnamefont {L.}~\bibnamefont {Miao}}, \bibinfo {author} {\bibfnamefont {Y.}~\bibnamefont {Lee}}, \bibinfo {author} {\bibfnamefont {A.}~\bibnamefont {Mei}}, \bibinfo {author} {\bibfnamefont {M.}~\bibnamefont {Lawler}},\ and\ \bibinfo {author} {\bibfnamefont {K.}~\bibnamefont {Shen}},\ }\bibfield  {title} {\bibinfo {title} {Two-dimensional magnetic monopole gas in an oxide heterostructure},\ }\href@noop {} {\bibfield  {journal} {\bibinfo  {journal} {Nature Communications}\ }\textbf {\bibinfo {volume} {11}},\ \bibinfo {pages} {1341} (\bibinfo {year} {2020})}\BibitemShut {NoStop}%
\bibitem [{\citenamefont {Lantagne-Hurtubise}\ \emph {et~al.}(2018)\citenamefont {Lantagne-Hurtubise}, \citenamefont {Rau},\ and\ \citenamefont {Gingras}}]{lantagne2018spin}%
  \BibitemOpen
  \bibfield  {author} {\bibinfo {author} {\bibfnamefont {{\'E}.}~\bibnamefont {Lantagne-Hurtubise}}, \bibinfo {author} {\bibfnamefont {J.~G.}\ \bibnamefont {Rau}},\ and\ \bibinfo {author} {\bibfnamefont {M.~J.}\ \bibnamefont {Gingras}},\ }\bibfield  {title} {\bibinfo {title} {Spin-ice thin films: large-n theory and monte carlo simulations},\ }\href@noop {} {\bibfield  {journal} {\bibinfo  {journal} {Physical Review X}\ }\textbf {\bibinfo {volume} {8}},\ \bibinfo {pages} {021053} (\bibinfo {year} {2018})}\BibitemShut {NoStop}%
\bibitem [{\citenamefont {Jaubert}\ \emph {et~al.}(2017)\citenamefont {Jaubert}, \citenamefont {Lin}, \citenamefont {Opel}, \citenamefont {Holdsworth},\ and\ \citenamefont {Gingras}}]{jaubert2017spin}%
  \BibitemOpen
  \bibfield  {author} {\bibinfo {author} {\bibfnamefont {L.}~\bibnamefont {Jaubert}}, \bibinfo {author} {\bibfnamefont {T.}~\bibnamefont {Lin}}, \bibinfo {author} {\bibfnamefont {T.}~\bibnamefont {Opel}}, \bibinfo {author} {\bibfnamefont {P.}~\bibnamefont {Holdsworth}},\ and\ \bibinfo {author} {\bibfnamefont {M.}~\bibnamefont {Gingras}},\ }\bibfield  {title} {\bibinfo {title} {Spin ice thin film: surface ordering, emergent square ice, and strain effects},\ }\href@noop {} {\bibfield  {journal} {\bibinfo  {journal} {Physical Review Letters}\ }\textbf {\bibinfo {volume} {118}},\ \bibinfo {pages} {207206} (\bibinfo {year} {2017})}\BibitemShut {NoStop}%
\bibitem [{\citenamefont {Zhang}\ \emph {et~al.}(2023)\citenamefont {Zhang}, \citenamefont {Xing}, \citenamefont {Noordhoek}, \citenamefont {Liu}, \citenamefont {Zhao}, \citenamefont {Hor{\'a}k}, \citenamefont {Huang}, \citenamefont {Hao}, \citenamefont {Yang}, \citenamefont {Pandey} \emph {et~al.}}]{zhang2023anomalous}%
  \BibitemOpen
  \bibfield  {author} {\bibinfo {author} {\bibfnamefont {H.}~\bibnamefont {Zhang}}, \bibinfo {author} {\bibfnamefont {C.}~\bibnamefont {Xing}}, \bibinfo {author} {\bibfnamefont {K.}~\bibnamefont {Noordhoek}}, \bibinfo {author} {\bibfnamefont {Z.}~\bibnamefont {Liu}}, \bibinfo {author} {\bibfnamefont {T.}~\bibnamefont {Zhao}}, \bibinfo {author} {\bibfnamefont {L.}~\bibnamefont {Hor{\'a}k}}, \bibinfo {author} {\bibfnamefont {Q.}~\bibnamefont {Huang}}, \bibinfo {author} {\bibfnamefont {L.}~\bibnamefont {Hao}}, \bibinfo {author} {\bibfnamefont {J.}~\bibnamefont {Yang}}, \bibinfo {author} {\bibfnamefont {S.}~\bibnamefont {Pandey}}, \emph {et~al.},\ }\bibfield  {title} {\bibinfo {title} {Anomalous magnetoresistance by breaking ice rule in \ch{Bi2Ir2O7/Dy2Ti2O7} heterostructure},\ }\href@noop {} {\bibfield  {journal} {\bibinfo  {journal} {Nature Communications}\ }\textbf {\bibinfo {volume} {14}},\ \bibinfo {pages} {1404} (\bibinfo {year} {2023})}\BibitemShut {NoStop}%
\bibitem [{\citenamefont {Qi}\ \emph {et~al.}(2012)\citenamefont {Qi}, \citenamefont {Korneta}, \citenamefont {Wan}, \citenamefont {DeLong}, \citenamefont {Schlottmann},\ and\ \citenamefont {Cao}}]{qi2012strong}%
  \BibitemOpen
  \bibfield  {author} {\bibinfo {author} {\bibfnamefont {T.}~\bibnamefont {Qi}}, \bibinfo {author} {\bibfnamefont {O.}~\bibnamefont {Korneta}}, \bibinfo {author} {\bibfnamefont {X.}~\bibnamefont {Wan}}, \bibinfo {author} {\bibfnamefont {L.}~\bibnamefont {DeLong}}, \bibinfo {author} {\bibfnamefont {P.}~\bibnamefont {Schlottmann}},\ and\ \bibinfo {author} {\bibfnamefont {G.}~\bibnamefont {Cao}},\ }\bibfield  {title} {\bibinfo {title} {Strong magnetic instability in correlated metallic \ch{Bi2Ir2O7}},\ }\href@noop {} {\bibfield  {journal} {\bibinfo  {journal} {Journal of Physics: Condensed Matter}\ }\textbf {\bibinfo {volume} {24}},\ \bibinfo {pages} {345601} (\bibinfo {year} {2012})}\BibitemShut {NoStop}%
\bibitem [{\citenamefont {Chu}\ \emph {et~al.}(2019)\citenamefont {Chu}, \citenamefont {Liu}, \citenamefont {Zhang}, \citenamefont {Noordhoek}, \citenamefont {Riggs}, \citenamefont {Shapiro}, \citenamefont {Serro}, \citenamefont {Yi}, \citenamefont {Mellisa}, \citenamefont {Suresha} \emph {et~al.}}]{chu2019possible}%
  \BibitemOpen
  \bibfield  {author} {\bibinfo {author} {\bibfnamefont {J.-H.}\ \bibnamefont {Chu}}, \bibinfo {author} {\bibfnamefont {J.}~\bibnamefont {Liu}}, \bibinfo {author} {\bibfnamefont {H.}~\bibnamefont {Zhang}}, \bibinfo {author} {\bibfnamefont {K.}~\bibnamefont {Noordhoek}}, \bibinfo {author} {\bibfnamefont {S.~C.}\ \bibnamefont {Riggs}}, \bibinfo {author} {\bibfnamefont {M.}~\bibnamefont {Shapiro}}, \bibinfo {author} {\bibfnamefont {C.~R.}\ \bibnamefont {Serro}}, \bibinfo {author} {\bibfnamefont {D.}~\bibnamefont {Yi}}, \bibinfo {author} {\bibfnamefont {M.}~\bibnamefont {Mellisa}}, \bibinfo {author} {\bibfnamefont {S.}~\bibnamefont {Suresha}}, \emph {et~al.},\ }\bibfield  {title} {\bibinfo {title} {Possible scale invariant linear magnetoresistance in pyrochlore iridates \ch{Bi2Ir2O7}},\ }\href@noop {} {\bibfield  {journal} {\bibinfo  {journal} {New Journal of Physics}\ }\textbf {\bibinfo {volume} {21}},\ \bibinfo {pages} {113041} (\bibinfo {year} {2019})}\BibitemShut {NoStop}%
\bibitem [{\citenamefont {Yang}\ \emph {et~al.}(2018)\citenamefont {Yang}, \citenamefont {Xie}, \citenamefont {Sun}, \citenamefont {Zhang}, \citenamefont {Park}, \citenamefont {Xue}, \citenamefont {Li}, \citenamefont {Tao}, \citenamefont {Jia}, \citenamefont {Losovyj} \emph {et~al.}}]{yang2018stoichiometry}%
  \BibitemOpen
  \bibfield  {author} {\bibinfo {author} {\bibfnamefont {W.}~\bibnamefont {Yang}}, \bibinfo {author} {\bibfnamefont {Y.}~\bibnamefont {Xie}}, \bibinfo {author} {\bibfnamefont {X.}~\bibnamefont {Sun}}, \bibinfo {author} {\bibfnamefont {X.}~\bibnamefont {Zhang}}, \bibinfo {author} {\bibfnamefont {K.}~\bibnamefont {Park}}, \bibinfo {author} {\bibfnamefont {S.}~\bibnamefont {Xue}}, \bibinfo {author} {\bibfnamefont {Y.}~\bibnamefont {Li}}, \bibinfo {author} {\bibfnamefont {C.}~\bibnamefont {Tao}}, \bibinfo {author} {\bibfnamefont {Q.}~\bibnamefont {Jia}}, \bibinfo {author} {\bibfnamefont {Y.}~\bibnamefont {Losovyj}}, \emph {et~al.},\ }\bibfield  {title} {\bibinfo {title} {Stoichiometry control and electronic and transport properties of pyrochlore \ch{Bi2Ir2O7} thin films},\ }\href@noop {} {\bibfield  {journal} {\bibinfo  {journal} {Physical Review Materials}\ }\textbf {\bibinfo {volume} {2}},\ \bibinfo {pages} {114206} (\bibinfo {year} {2018})}\BibitemShut {NoStop}%
\bibitem [{\citenamefont {Sato}\ \emph {et~al.}(2007)\citenamefont {Sato}, \citenamefont {Matsuhira}, \citenamefont {Sakakibara}, \citenamefont {Tayama}, \citenamefont {Hiroi},\ and\ \citenamefont {Takagi}}]{sato2007field}%
  \BibitemOpen
  \bibfield  {author} {\bibinfo {author} {\bibfnamefont {H.}~\bibnamefont {Sato}}, \bibinfo {author} {\bibfnamefont {K.}~\bibnamefont {Matsuhira}}, \bibinfo {author} {\bibfnamefont {T.}~\bibnamefont {Sakakibara}}, \bibinfo {author} {\bibfnamefont {T.}~\bibnamefont {Tayama}}, \bibinfo {author} {\bibfnamefont {Z.}~\bibnamefont {Hiroi}},\ and\ \bibinfo {author} {\bibfnamefont {S.}~\bibnamefont {Takagi}},\ }\bibfield  {title} {\bibinfo {title} {Field-angle dependence of the ice-rule breaking spin-flip transition in \ch{Dy2Ti2O7}},\ }\href@noop {} {\bibfield  {journal} {\bibinfo  {journal} {Journal of Physics: Condensed Matter}\ }\textbf {\bibinfo {volume} {19}},\ \bibinfo {pages} {145272} (\bibinfo {year} {2007})}\BibitemShut {NoStop}%
\bibitem [{\citenamefont {Ohno}\ \emph {et~al.}(2024)\citenamefont {Ohno}, \citenamefont {Fujita},\ and\ \citenamefont {Kawasaki}}]{ohno2024proximity}%
  \BibitemOpen
  \bibfield  {author} {\bibinfo {author} {\bibfnamefont {M.}~\bibnamefont {Ohno}}, \bibinfo {author} {\bibfnamefont {T.~C.}\ \bibnamefont {Fujita}},\ and\ \bibinfo {author} {\bibfnamefont {M.}~\bibnamefont {Kawasaki}},\ }\bibfield  {title} {\bibinfo {title} {Proximity effect of emergent field from spin ice in an oxide heterostructure},\ }\href@noop {} {\bibfield  {journal} {\bibinfo  {journal} {Science Advances}\ }\textbf {\bibinfo {volume} {10}},\ \bibinfo {pages} {eadk6308} (\bibinfo {year} {2024})}\BibitemShut {NoStop}%
\bibitem [{\citenamefont {Li}(2008)}]{li2008torque}%
  \BibitemOpen
  \bibfield  {author} {\bibinfo {author} {\bibfnamefont {L.}~\bibnamefont {Li}},\ }\href@noop {} {\emph {\bibinfo {title} {Torque magnetometry in unconventional superconductors}}}\ (\bibinfo  {publisher} {Princeton University},\ \bibinfo {year} {2008})\BibitemShut {NoStop}%
\bibitem [{\citenamefont {Anand}\ \emph {et~al.}(2022)\citenamefont {Anand}, \citenamefont {Barry}, \citenamefont {Neu}, \citenamefont {Graf}, \citenamefont {Huang}, \citenamefont {Zhou}, \citenamefont {Siegrist}, \citenamefont {Changlani},\ and\ \citenamefont {Beekman}}]{anand2022investigation}%
  \BibitemOpen
  \bibfield  {author} {\bibinfo {author} {\bibfnamefont {N.}~\bibnamefont {Anand}}, \bibinfo {author} {\bibfnamefont {K.}~\bibnamefont {Barry}}, \bibinfo {author} {\bibfnamefont {J.~N.}\ \bibnamefont {Neu}}, \bibinfo {author} {\bibfnamefont {D.~E.}\ \bibnamefont {Graf}}, \bibinfo {author} {\bibfnamefont {Q.}~\bibnamefont {Huang}}, \bibinfo {author} {\bibfnamefont {H.}~\bibnamefont {Zhou}}, \bibinfo {author} {\bibfnamefont {T.}~\bibnamefont {Siegrist}}, \bibinfo {author} {\bibfnamefont {H.~J.}\ \bibnamefont {Changlani}},\ and\ \bibinfo {author} {\bibfnamefont {C.}~\bibnamefont {Beekman}},\ }\bibfield  {title} {\bibinfo {title} {Investigation of the monopole magneto-chemical potential in spin ices using capacitive torque magnetometry},\ }\href@noop {} {\bibfield  {journal} {\bibinfo  {journal} {Nature Communications}\ }\textbf {\bibinfo {volume} {13}},\ \bibinfo {pages} {3818} (\bibinfo {year} {2022})}\BibitemShut {NoStop}%
\bibitem [{\citenamefont {Borzi}\ \emph {et~al.}(2016)\citenamefont {Borzi}, \citenamefont {G{\'o}mez~Albarrac{\'\i}n}, \citenamefont {Rosales}, \citenamefont {Rossini}, \citenamefont {Steppke}, \citenamefont {Prabhakaran}, \citenamefont {Mackenzie}, \citenamefont {Cabra},\ and\ \citenamefont {Grigera}}]{borzi2016intermediate}%
  \BibitemOpen
  \bibfield  {author} {\bibinfo {author} {\bibfnamefont {R.~A.}\ \bibnamefont {Borzi}}, \bibinfo {author} {\bibfnamefont {F.~A.}\ \bibnamefont {G{\'o}mez~Albarrac{\'\i}n}}, \bibinfo {author} {\bibfnamefont {H.~D.}\ \bibnamefont {Rosales}}, \bibinfo {author} {\bibfnamefont {G.~L.}\ \bibnamefont {Rossini}}, \bibinfo {author} {\bibfnamefont {A.}~\bibnamefont {Steppke}}, \bibinfo {author} {\bibfnamefont {D.}~\bibnamefont {Prabhakaran}}, \bibinfo {author} {\bibfnamefont {A.}~\bibnamefont {Mackenzie}}, \bibinfo {author} {\bibfnamefont {D.~C.}\ \bibnamefont {Cabra}},\ and\ \bibinfo {author} {\bibfnamefont {S.~A.}\ \bibnamefont {Grigera}},\ }\bibfield  {title} {\bibinfo {title} {Intermediate magnetization state and competing orders in \ch{Dy2Ti2O7} and \ch{Ho2Ti2O7}},\ }\href@noop {} {\bibfield  {journal} {\bibinfo  {journal} {Nature Communications}\ }\textbf {\bibinfo {volume} {7}},\ \bibinfo {pages} {12592} (\bibinfo {year} {2016})}\BibitemShut {NoStop}%
\bibitem [{sup()}]{supplemental}%
  \BibitemOpen
  \href@noop {} {\bibinfo {title} {See {S}upplemental {M}aterial at [url] for [heterostructure synthesis; {X}-ray characterization; transmission electron microscopes; resistivity and torque measurements; fitting methods; measurement reproducibility; consideration of potential inhomogeneity; temperature-dependent resistivity from 300 {K} to 3.7 {K}], which includes refs. [56]}}\BibitemShut {NoStop}%
\bibitem [{\citenamefont {Zhou}\ \emph {et~al.}(2012)\citenamefont {Zhou}, \citenamefont {Cheng}, \citenamefont {Hallas}, \citenamefont {Wiebe}, \citenamefont {Li}, \citenamefont {Balicas}, \citenamefont {Zhou}, \citenamefont {Goodenough}, \citenamefont {Gardner},\ and\ \citenamefont {Choi}}]{zhou2012chemical}%
  \BibitemOpen
  \bibfield  {author} {\bibinfo {author} {\bibfnamefont {H.}~\bibnamefont {Zhou}}, \bibinfo {author} {\bibfnamefont {J.}~\bibnamefont {Cheng}}, \bibinfo {author} {\bibfnamefont {A.}~\bibnamefont {Hallas}}, \bibinfo {author} {\bibfnamefont {C.}~\bibnamefont {Wiebe}}, \bibinfo {author} {\bibfnamefont {G.}~\bibnamefont {Li}}, \bibinfo {author} {\bibfnamefont {L.}~\bibnamefont {Balicas}}, \bibinfo {author} {\bibfnamefont {J.}~\bibnamefont {Zhou}}, \bibinfo {author} {\bibfnamefont {J.}~\bibnamefont {Goodenough}}, \bibinfo {author} {\bibfnamefont {J.~S.}\ \bibnamefont {Gardner}},\ and\ \bibinfo {author} {\bibfnamefont {E.}~\bibnamefont {Choi}},\ }\bibfield  {title} {\bibinfo {title} {Chemical pressure effects on pyrochlore spin ice},\ }\href@noop {} {\bibfield  {journal} {\bibinfo  {journal} {Physical Review Letters}\ }\textbf {\bibinfo {volume} {108}},\ \bibinfo {pages} {207206} (\bibinfo {year} {2012})}\BibitemShut {NoStop}%
\bibitem [{\citenamefont {Ruminy}\ \emph {et~al.}(2016)\citenamefont {Ruminy}, \citenamefont {Pomjakushina}, \citenamefont {Iida}, \citenamefont {Kamazawa}, \citenamefont {Adroja}, \citenamefont {Stuhr},\ and\ \citenamefont {Fennell}}]{ruminy2016crystal}%
  \BibitemOpen
  \bibfield  {author} {\bibinfo {author} {\bibfnamefont {M.}~\bibnamefont {Ruminy}}, \bibinfo {author} {\bibfnamefont {E.}~\bibnamefont {Pomjakushina}}, \bibinfo {author} {\bibfnamefont {K.}~\bibnamefont {Iida}}, \bibinfo {author} {\bibfnamefont {K.}~\bibnamefont {Kamazawa}}, \bibinfo {author} {\bibfnamefont {D.}~\bibnamefont {Adroja}}, \bibinfo {author} {\bibfnamefont {U.}~\bibnamefont {Stuhr}},\ and\ \bibinfo {author} {\bibfnamefont {T.}~\bibnamefont {Fennell}},\ }\bibfield  {title} {\bibinfo {title} {Crystal-field parameters of the rare-earth pyrochlores r 2 ti 2 o 7 (r= tb, dy, and ho)},\ }\href@noop {} {\bibfield  {journal} {\bibinfo  {journal} {Physical Review B}\ }\textbf {\bibinfo {volume} {94}},\ \bibinfo {pages} {024430} (\bibinfo {year} {2016})}\BibitemShut {NoStop}%
\bibitem [{\citenamefont {Lu}\ \emph {et~al.}(2024)\citenamefont {Lu}, \citenamefont {Sch{\"a}fer}, \citenamefont {Hall{\'e}n},\ and\ \citenamefont {Laumann}}]{lu2024111}%
  \BibitemOpen
  \bibfield  {author} {\bibinfo {author} {\bibfnamefont {Z.}~\bibnamefont {Lu}}, \bibinfo {author} {\bibfnamefont {R.}~\bibnamefont {Sch{\"a}fer}}, \bibinfo {author} {\bibfnamefont {J.~N.}\ \bibnamefont {Hall{\'e}n}},\ and\ \bibinfo {author} {\bibfnamefont {C.~R.}\ \bibnamefont {Laumann}},\ }\bibfield  {title} {\bibinfo {title} {[111]-strained spin ice: Localization of thermodynamically deconfined monopoles},\ }\href@noop {} {\bibfield  {journal} {\bibinfo  {journal} {Physical Review B}\ }\textbf {\bibinfo {volume} {110}},\ \bibinfo {pages} {184421} (\bibinfo {year} {2024})}\BibitemShut {NoStop}%
\bibitem [{\citenamefont {Xing}\ \emph {et~al.}(2024)\citenamefont {Xing}, \citenamefont {Zhang}, \citenamefont {Yao}, \citenamefont {Cui}, \citenamefont {Huang}, \citenamefont {Yang}, \citenamefont {Pandey}, \citenamefont {Gong}, \citenamefont {Hor{\'a}k}, \citenamefont {Xin} \emph {et~al.}}]{xing2024anomalous}%
  \BibitemOpen
  \bibfield  {author} {\bibinfo {author} {\bibfnamefont {C.}~\bibnamefont {Xing}}, \bibinfo {author} {\bibfnamefont {S.}~\bibnamefont {Zhang}}, \bibinfo {author} {\bibfnamefont {W.}~\bibnamefont {Yao}}, \bibinfo {author} {\bibfnamefont {D.}~\bibnamefont {Cui}}, \bibinfo {author} {\bibfnamefont {Q.}~\bibnamefont {Huang}}, \bibinfo {author} {\bibfnamefont {J.}~\bibnamefont {Yang}}, \bibinfo {author} {\bibfnamefont {S.}~\bibnamefont {Pandey}}, \bibinfo {author} {\bibfnamefont {D.}~\bibnamefont {Gong}}, \bibinfo {author} {\bibfnamefont {L.}~\bibnamefont {Hor{\'a}k}}, \bibinfo {author} {\bibfnamefont {Y.}~\bibnamefont {Xin}}, \emph {et~al.},\ }\bibfield  {title} {\bibinfo {title} {Anomalous proximitized transport in metal/quantum magnet heterostructure \ch{Bi2Ir2O7/Yb2Ti2O7}},\ }\href@noop {} {\bibfield  {journal} {\bibinfo  {journal} {Physical Review Materials}\ }\textbf {\bibinfo {volume} {8}},\ \bibinfo {pages} {114407} (\bibinfo {year} {2024})}\BibitemShut {NoStop}%
\bibitem [{\citenamefont {Bovo}\ \emph {et~al.}(2016)\citenamefont {Bovo}, \citenamefont {Rouleau}, \citenamefont {Prabhakaran},\ and\ \citenamefont {Bramwell}}]{bovo2016layer}%
  \BibitemOpen
  \bibfield  {author} {\bibinfo {author} {\bibfnamefont {L.}~\bibnamefont {Bovo}}, \bibinfo {author} {\bibfnamefont {C.~M.}\ \bibnamefont {Rouleau}}, \bibinfo {author} {\bibfnamefont {D.}~\bibnamefont {Prabhakaran}},\ and\ \bibinfo {author} {\bibfnamefont {S.~T.}\ \bibnamefont {Bramwell}},\ }\bibfield  {title} {\bibinfo {title} {Layer-by-layer epitaxial thin films of the pyrochlore \ch{Tb2Ti2O7}},\ }\href@noop {} {\bibfield  {journal} {\bibinfo  {journal} {Nanotechnology}\ }\textbf {\bibinfo {volume} {28}},\ \bibinfo {pages} {055708} (\bibinfo {year} {2016})}\BibitemShut {NoStop}%
\bibitem [{\citenamefont {Wu}\ \emph {et~al.}(2024)\citenamefont {Wu}, \citenamefont {Chang}, \citenamefont {Wu}, \citenamefont {Terilli}, \citenamefont {Wen}, \citenamefont {Kareev}, \citenamefont {Choi}, \citenamefont {Graf}, \citenamefont {Zhang}, \citenamefont {Gu} \emph {et~al.}}]{wu2024electronic}%
  \BibitemOpen
  \bibfield  {author} {\bibinfo {author} {\bibfnamefont {T.-C.}\ \bibnamefont {Wu}}, \bibinfo {author} {\bibfnamefont {Y.}~\bibnamefont {Chang}}, \bibinfo {author} {\bibfnamefont {A.-K.}\ \bibnamefont {Wu}}, \bibinfo {author} {\bibfnamefont {M.}~\bibnamefont {Terilli}}, \bibinfo {author} {\bibfnamefont {F.}~\bibnamefont {Wen}}, \bibinfo {author} {\bibfnamefont {M.}~\bibnamefont {Kareev}}, \bibinfo {author} {\bibfnamefont {E.~S.}\ \bibnamefont {Choi}}, \bibinfo {author} {\bibfnamefont {D.}~\bibnamefont {Graf}}, \bibinfo {author} {\bibfnamefont {Q.}~\bibnamefont {Zhang}}, \bibinfo {author} {\bibfnamefont {L.}~\bibnamefont {Gu}}, \emph {et~al.},\ }\bibfield  {title} {\bibinfo {title} {Electronic anisotropy and rotational symmetry breaking at a weyl semimetal/spin ice interface},\ }\href@noop {} {\bibfield  {journal} {\bibinfo  {journal} {arXiv preprint arXiv:2409.18880}\ } (\bibinfo {year} {2024})}\BibitemShut {NoStop}%
\bibitem [{\citenamefont {Kareev}\ \emph {et~al.}(2025)\citenamefont {Kareev}, \citenamefont {Liu}, \citenamefont {Terilli}, \citenamefont {Wen}, \citenamefont {Wu}, \citenamefont {Doughty}, \citenamefont {Li}, \citenamefont {Zhou}, \citenamefont {Zhang}, \citenamefont {Gu} \emph {et~al.}}]{kareev2025epitaxial}%
  \BibitemOpen
  \bibfield  {author} {\bibinfo {author} {\bibfnamefont {M.}~\bibnamefont {Kareev}}, \bibinfo {author} {\bibfnamefont {X.}~\bibnamefont {Liu}}, \bibinfo {author} {\bibfnamefont {M.}~\bibnamefont {Terilli}}, \bibinfo {author} {\bibfnamefont {F.}~\bibnamefont {Wen}}, \bibinfo {author} {\bibfnamefont {T.-C.}\ \bibnamefont {Wu}}, \bibinfo {author} {\bibfnamefont {D.}~\bibnamefont {Doughty}}, \bibinfo {author} {\bibfnamefont {H.}~\bibnamefont {Li}}, \bibinfo {author} {\bibfnamefont {J.}~\bibnamefont {Zhou}}, \bibinfo {author} {\bibfnamefont {Q.}~\bibnamefont {Zhang}}, \bibinfo {author} {\bibfnamefont {L.}~\bibnamefont {Gu}}, \emph {et~al.},\ }\bibfield  {title} {\bibinfo {title} {Epitaxial stabilization of a pyrochlore interface between weyl semimetal and spin ice},\ }\href@noop {} {\bibfield  {journal} {\bibinfo  {journal} {Nano Letters}\ } (\bibinfo {year} {2025})}\BibitemShut {NoStop}%
\bibitem [{\citenamefont {Yang}\ \emph {et~al.}(2017)\citenamefont {Yang}, \citenamefont {Xie}, \citenamefont {Zhu}, \citenamefont {Park}, \citenamefont {Chen}, \citenamefont {Losovyj}, \citenamefont {Li}, \citenamefont {Liu}, \citenamefont {Starr}, \citenamefont {Acosta} \emph {et~al.}}]{yang2017epitaxial}%
  \BibitemOpen
  \bibfield  {author} {\bibinfo {author} {\bibfnamefont {W.}~\bibnamefont {Yang}}, \bibinfo {author} {\bibfnamefont {Y.}~\bibnamefont {Xie}}, \bibinfo {author} {\bibfnamefont {W.}~\bibnamefont {Zhu}}, \bibinfo {author} {\bibfnamefont {K.}~\bibnamefont {Park}}, \bibinfo {author} {\bibfnamefont {A.}~\bibnamefont {Chen}}, \bibinfo {author} {\bibfnamefont {Y.}~\bibnamefont {Losovyj}}, \bibinfo {author} {\bibfnamefont {Z.}~\bibnamefont {Li}}, \bibinfo {author} {\bibfnamefont {H.}~\bibnamefont {Liu}}, \bibinfo {author} {\bibfnamefont {M.}~\bibnamefont {Starr}}, \bibinfo {author} {\bibfnamefont {J.~A.}\ \bibnamefont {Acosta}}, \emph {et~al.},\ }\bibfield  {title} {\bibinfo {title} {Epitaxial thin films of pyrochlore iridate \ch{Bi_{2+x}Ir_{2-y}O_{7-$\delta$}}: Structure, defects and transport properties},\ }\href@noop {} {\bibfield  {journal} {\bibinfo  {journal} {Scientific Reports}\ }\textbf {\bibinfo {volume} {7}},\ \bibinfo {pages} {7740} (\bibinfo {year} {2017})}\BibitemShut {NoStop}%
\end{thebibliography}%
    \section*{End Matter}
    \addcontentsline{toc}{section}{End Matter}
    \subsection{Experimental Methods}
    We synthesized the epitaxial DTO thin film of $\sim$18 nm thickness on the (111)-oriented Yttria Stabilized Zirconia (YSZ) substrate, which is the most common and commercially available substrate for pyrochlore thin films. This thickness was chosen so that a significant bulk region still exists within the film without significant relaxation. The DTO film was capped by an epitaxial BIO thin film of $\sim$3 nm thickness [Fig. \ref{fig1}(a)] to enable proximitized transport measurements. The BIO thin film thickness was chosen to be similar to the previous work \cite{zhang2023anomalous} \textcolor{black}{which already demonstrated the proximitized transport behavior is weaker when BIO is thick}. Most details of the growth are discussed in the supplement \cite{supplemental}. Scanning transmission electron microscopy (STEM) images in Fig. S4 \cite{supplemental} showed a highly epitaxial pyrochlore structure for both the DTO and BIO layers as well as sharp BIO/DTO and DTO/YSZ interfaces. No significant interdiffusion across the interface was found by energy-dispersive X-ray spectroscopic (EDS) map [Fig. S5] \cite{supplemental}. The epitaxial growth of both DTO and BIO layers was also confirmed by the specular scan of X-ray diffraction [Fig. S6(a)] \cite{supplemental}. Reciprocal Space Mapping (RSM) [Fig. S6(b)] \cite{supplemental} showed that the out-of-plane lattice spacing $d_{111}$ of the DTO film is 5.75 $\rm{\AA}$ \textcolor{black}{$\pm$ 0.001 $\rm{\AA}$}, which is roughly 1.7\% smaller than bulk DTO, and the in-plane spacing $d_{11-2}$ is 4.19 $\rm{\AA}$ \textcolor{black}{$\pm$ 0.01 $\rm{\AA}$}, which is 1.4\% larger than bulk DTO. That corresponds to a volume expansion of about 1\%. The BIO reflection was not resolved within the measurement statistics of in-house RSM likely due to the small thickness. From X-ray diffraction [Fig. S6(a)] \cite{supplemental} the out-of-plane lattice spacing $d_{111}$ of the BIO film is 6.10 
    $\rm{\AA}$, which is 2.5\% larger than bulk BIO. The thickness of the BIO layer and DTO layer is verified by X-ray reflectivity [Fig. S6(c)] \cite{supplemental}.

\end{document}